\documentclass[12pt,preprint]{aastex}

\shorttitle{SPECTROPOLARIMETRY OF THE Ba~{$\!$\sc ii} D LINES}
\shortauthors{BELLUZZI, TRUJILLO BUENO \& LANDI DEGL'INNOCENTI}

\begin{document}

\title{THE MAGNETIC SENSITIVITY OF THE Ba~{$\!$\sc ii} D$_1$ AND D$_2$\\ 
LINES OF THE FRAUNHOFER SPECTRUM}

\author{L{\sc{uca}} B{\sc{elluzzi}}\altaffilmark{1,2}, 
J{\sc{avier}} T{\sc{rujillo}} B{\sc{ueno}}\altaffilmark{1,3} \\
{\sc{and}} \\ 
E{\sc{gidio}} L{\sc{andi}} D{\sc{egl}}'I{\sc{nnocenti}}\altaffilmark{2}}

\altaffiltext{1}{Instituto de Astrof\'\i{}sica de Canarias, V\'\i{}a L\'actea 
s/n, E-38205 La Laguna, Tenerife, Spain}
\altaffiltext{2}{Universit\`a degli Studi di Firenze, Dipartimento di 
Astronomia e Scienza dello Spazio, \\
Largo Enrico Fermi 2, I-50125 Firenze, Italy}
\altaffiltext{3}{Consejo Superior de Investigaciones Cient\'\i ficas, Spain}

\begin{abstract}
The physical interpretation of the spectral line polarization produced by the
joint action of the Hanle and Zeeman effects offers a unique opportunity to 
obtain empirical information about hidden aspects of solar and stellar 
magnetism. To this end, it is important to achieve a complete understanding 
of the sensitivity of the emergent spectral line polarization to the presence
of a magnetic field. Here we present a detailed 
theoretical investigation on the role of resonance scattering and 
magnetic fields on the polarization signals of the Ba~{$\!$\sc ii} D$_1$ 
and D$_2$ lines of the Fraunhofer spectrum, respectively at 4934~{\AA} 
and 4554~{\AA}.
We adopt a three-level model of Ba~{$\!$\sc ii}, and we take into account the 
hyperfine structure that is shown by the $^{135}$Ba 
and $^{137}$Ba isotopes. Despite of their relatively small abundance (18\%), 
the contribution coming from these two isotopes is indeed fundamental for the 
interpretation of the polarization signals observed in these lines.
We consider an optically thin slab model, through which we can investigate in a rigorous way the essential physical mechanisms involved (resonance 
polarization, Zeeman, Paschen-Back and Hanle effects), avoiding complications 
due to radiative transfer effects. We assume the slab to be illuminated from 
below by the photospheric solar continuum radiation field, and we investigate 
the radiation scattered at 90$^\circ$, both in the absence and in 
the presence of magnetic fields, deterministic and microturbulent.
We show in particular the existence of a differential magnetic
sensitivity of the three-peak $Q/I$ profile that is observed in the D$_{2}$ 
line in quiet regions close to the solar limb, which is of great interest for 
magnetic field diagnostics.
\end{abstract}


\section{INTRODUCTION} 
\label{sect:introduction}
Probably, the most interesting aspect of spectropolarimetry is that
it allows us to diagnose magnetic fields in astrophysics. To this
end, it is crucial to achieve a complete physical understanding of
the magnetic sensitivity of the emergent spectral line radiation
given the fact that it can occur through a variety of rather
unfamiliar physical mechanisms, not only via the Zeeman effect (e.g.,
Landi Degl'Innocenti \& Landolfi 2004 (hereafter LL04); see also the 
reviews by Stenflo 2003 and Trujillo Bueno 2003). In this respect, the 
main aim of this paper is to help decipher the physical mechanisms that
control the magnetic sensitivity of the polarization of the D-lines
of Ba~{$\!$\sc ii}, with particular interest in developing a powerful
diagnostic tool for mapping the magnetic field of the lower solar
chromosphere.\\
\indent In the atmospheres of the Sun and of other stars there is a
fundamental mechanism producing polarization in spectral lines, which
has nothing to do with the familiar Zeeman effect. There, where light
escapes through the stellar ``surface'', the atomic system is
illuminated anisotropically. The radiative transitions
produce population imbalances and quantum coherences between pairs of
magnetic sublevels, even among those pertaining to different levels.
The mere presence of this so-called atomic level polarization
produces spectral line polarization, without the need of a magnetic field.
This is usually referred to as resonance line polarization.
The important point is that a magnetic field
can modify the atomic polarization of the upper and/or lower levels
of the spectral line under consideration and the ensuing polarization
of the emergent spectral line radiation. Interestingly, the possible
presence of crossings and repulsions among magnetic sublevels of 
fine-structure and/or hyperfine-structure multiplets can enhance
dramatically the magnetic sensitivity of the emergent spectral line
polarization. A remarkable example is the enhancement of the line-core 
scattering polarization of the D$_2$ line of Na~{$\!$\sc i} by a
vertical magnetic field, which is due to interferences between particular
hyperfine structure (HFS) magnetic sublevels of the $^2P_{3/2}$ upper level
\citep{jtb02}. It is of interest to note that this theoretical prediction
was observationally confirmed by \citet{Ste02} via filter
polarimetry of the solar atmosphere.\\
\indent In contrast with the case of sodium, which has one single isotope with 
nuclear spin I=3/2, barium has five (even) isotopes with I=0 (with an overall
abundance of $82.18\%$) and two (odd) isotopes with I=3/2 (with an
abundance of $17.82\%$).
Moreover, the HFS splitting of the odd isotopes of barium 
is about a factor five larger than for the case of sodium. 
Obviously, the emergent fractional linear polarization
(i.e., the $Q/I$ profile, where $I$ and $Q$ are two of the Stokes
parameters) has contributions from all the barium isotopes.
In fact, as pointed out by \citet{Ste97b}, the $Q/I$ pattern of the
Ba~{$\!$\sc ii} D$_2$ line that \citet{Ste97a} observed in very
quiet regions close to the solar limb shows a three-peak structure,
with a prominent central $Q/I$ peak due to the even isotopes (which
are devoid of HFS) and two less significant peaks in the red and blue
wings caused by the contributions from the odd isotopes (which have
HFS). Therefore, we think that for the D$_2$ line of Ba~{$\!$\sc ii} we
should also have enhancement of scattering polarization by a vertical
field, but only around such wing wavelengths because the required
interferences occur only between the magnetic sublevels of the 
$^2P_{3/2}$ upper level of the barium isotopes endowed of HFS.
Actually, the scientific motivation that led us to undertake the
theoretical investigation presented here was to develop a novel
plasma diagnostic tool based on the idea that such isotopes of barium
must have a different behavior in the presence of a magnetic field, 
with respect to those devoid of HFS.\\
\indent While the physical origin of the observed $Q/I$ profile of the 
Ba~{$\!$\sc ii} D$_2$ line seems to be clear, nobody has yet been able to
model the $Q/I$ profiles
of the Ba~{$\!$\sc ii} D$_1$ line that \citet{Ste98} observed
in two different regions close to the solar limb. Interestingly, while
the $Q/I$ profile shown in Fig.~1 of \citet{Ste98} might
perhaps be the result of the interaction of the D$_1$ line with the
continuum (see the bottom panel to the right-hand-side), the
symmetric $Q/I$ profile shown in the panel just above the previous
one is very similar to the enigmatic $Q/I$ profile of the Na~{$\!$\sc i} 
D$_1$ line \citep[see also Fig.~3 in][]{Ste00}. Although the main
scientific target of our paper is to understand the magnetic
sensitivity of the above-mentioned Ba~{$\!$\sc ii} D$_2$ line, we present
also some results of our $Q/I$ calculations for the D$_1$ line with
the aim of helping to clarify its physical origin.\\
\indent The outline of this paper is the following. The formulation of the
problem is presented in Section~\ref{sect:formulation}, where we establish 
our modeling assumptions, we briefly discuss the relevant equations, 
and we describe the atomic model, showing the behaviour of the magnetic 
sublevels of the odd barium isotopes in the presence of an increasing 
magnetic field. The magnetic sensitivity of the atomic
polarization of the lower and upper levels of the Ba~{$\!$\sc ii} D-lines
is discussed in Section~\ref{sect:polarization}, pointing out the 
similarities with the behavior of the Na~{$\!$\sc i} levels. 
Section~\ref{sect:emergent} focuses on the emergent spectral line
polarization in the absence and in the presence of a 
magnetic field, with emphasis on the D$_2$ line of Ba~{$\!$\sc ii}, but showing 
also some interesting results for the D$_1$ line. Finally, 
Section~\ref{sect:conclusions} summarizes
our main conclusions with an outlook to future research. The three
appendices give detailed information that may help the reader to
understand better the complexity of the problem we are investigating.

\section{FORMULATION OF THE PROBLEM} 
\label{sect:formulation}
\begin{figure}[!t]
\begin{center} 
\includegraphics[width=0.5\textwidth]{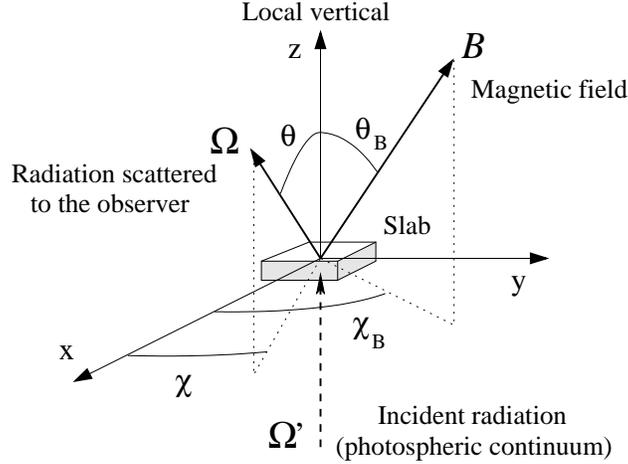}
\caption{\footnotesize{Geometry of the problem under investigation.}}
\label{fig:geometry}
\end{center}
\end{figure}

We consider an optically thin slab of Ba~{$\!$\sc ii} ions illuminated 
from below by the photospheric solar continuum radiation field
(see Fig.~\ref{fig:geometry}). 
The slab is assumed to be located 1000~km above the $\tau_{5000}\!=\!1$ 
photospheric level, the approximate height at which, according to 
semi-empirical models of the solar atmosphere, the line-core optical depth of 
the D$_2$ line is unity along a line-of-sight specified by 
$\mu\!\equiv\!{\rm cos}{\theta}\!=\!0.1$, with ${\theta}$ the 
heliocentric angle.\\
\indent In the following subsections we summarize the basic
equations derived within the framework of the quantum theory of spectral
line polarization that will be used
in this work (\S~\ref{sect:equations}), we describe the atomic model
that we have adopted to describe the Ba~{$\!$\sc ii} ion (\S~\ref{sect:atom}),
and finally we show how the main properties of the continuum photospheric
radiation field incident on the slab have been calculated
(\S~\ref{sect:radiation}).

\subsection{{\it The Basic Equations}}
\label{sect:equations}
We describe the excitation state of the Ba~{$\!$\sc ii} levels using 
the density matrix formalism, a robust theoretical framework very suitable to 
treat the atomic polarization (population imbalances and quantum interferences,
or coherences, among the magnetic sublevels) that anisotropic pumping processes 
induce in an atomic system. 
Referring to one of the isotopes of barium (see \S~\ref{sect:atom}) that 
shows HFS, we indicate with $I$ its nuclear
spin quantum number. In the absence of magnetic fields, using Dirac's 
notation, the energy eigenvectors can be written in the form 
$|\alpha \, J \, I \, F \, f >$,
where $\alpha$ represents a set of inner quantum numbers
(specifying the electronic configuration and, if the atomic system is described 
in the L-S coupling scheme, the total electronic orbital and spin angular 
momenta), $J$ is the total electronic angular momentum quantum number, 
$F$ is the total angular momentum quantum number (electronic plus nuclear:
{\boldmath $F$=$J$+$I$}), and $f$ is its projection along the 
quantization axis.
In principle, to have a suitable description of atomic polarization, one has 
to take into account all the coherences of the form
\begin{equation}
<\alpha\,J\,I\,F\,f\,|\,\hat{\rho}\,|\,\alpha\,J^{\prime}\,I\,
F^{\prime}\,f^{\prime}> \;\;,
\end{equation}
where $\hat{\rho}$ is the density matrix operator.
The approximation that is used in this paper 
consists in restricting the description of the atomic system to the 
$J$-diagonal density matrix elements
\begin{equation}
<\alpha\,J\,I\,F\,f\,|\,\hat{\rho}\,|\,\alpha\,J\,I\,F^{\prime}\,f^{\prime}> \;\;,
\end{equation}
or, in other words, in neglecting coherences between different $J$-levels.
The resulting model-atom is referred to as the ``multi-level atom with 
hyperfine structure'' according to LL04.
The approximation appears to be fully justified for the investigation of the 
Ba~{$\!$\sc ii} D-lines given the large frequency separation between the levels 
$6p\;^{2}P_{1/2}$ and $6p\;^{2}P_{3/2}$ (see \S~\ref{sect:atom}), and given the 
relatively low abundance of barium in the solar 
atmosphere\footnote{Note that this 
approximation is not justified for the D-lines of Na~{$\!$\sc i} and for the 
H and K lines of Ca~{$\!$\sc ii}, given the larger abundance of these ions.}.\\
\indent In the presence of a magnetic field, according to the general approach 
of the Paschen-Back effect theory, the energy eigenvalues and eigenvectors have 
to be found by diagonalization of the total
Hamiltonian (unperturbed atomic Hamiltonian plus magnetic Hamiltonian) on
each $J$-level subspace.
Taking the quantization axis along the magnetic field direction, it can be 
demonstrated that
the total Hamiltonian commutes with the projection along the quantization
axis of the total angular momentum operator ($f$ is a good quantum
number), while, in general, it does not commute with the total angular
momentum operator. The eigenvectors of the total Hamiltonian can be expressed 
in the form (e.g. LL04)
\begin{equation}
|\alpha\,J\,I\,i\,f>=\sum_{F}C_{F}^{\,i}\,(\alpha\,J\,I,\,f\,)\,|\alpha\,
J\,I\,F\,f> \;\; ,
\label{eq:eigenvectors}
\end{equation}
where the index $i$ labels the energy eigenstates belonging to the
subspace corresponding to assigned values of the quantum numbers $\alpha$, $J$, 
$I$ and $f$, and where the coefficients $C_{F}^{\,i}$ can be chosen to be real.
In the energy eigenvectors representation the atomic system
will therefore be described by means of the matrix elements
\begin{equation}
<\alpha\,J\,I\,i\,f\,|\,\hat{\rho}\,|\alpha\,J\,I\,i^{\prime}\,f^{\prime}>
\equiv\rho_{\alpha\,J\,I}(i\,f,\,i^{\prime}\,f^{\prime}) \;\;.
\label{eq:rho}
\end{equation}
If the magnetic field is so weak that the magnetic energy is much smaller than 
the energy intervals between the HFS $F$-levels, we are in the so-called 
{\it Zeeman effect regime} (of HFS), where the energy eigenvectors are still 
of the form $|\alpha\, J\, I\, F\, f> $ ($C_{F}^{\,i} \simeq \delta_{Fi}$), 
and the splitting between the HFS magnetic sublevels is 
linear with the magnetic field strength.
For stronger magnetic fields it is necessary to apply the Paschen-Back effect 
theory, and one enters the so-called {\it incomplete Paschen-Back effect 
regime}.
In this regime the energy eigenvectors have the general form of
equation~(\ref{eq:eigenvectors}) (the magnetic field produces a $F$-mixing of 
the various HFS levels originating from a particular $J$-level) and, 
as we will show in detail for the case of Ba~{$\!$\sc ii} in 
\S~{\ref{sect:atom}},
the splitting among the various HFS magnetic sublevels is no longer linear with 
the magnetic field. Several crossings 
among HFS magnetic sublevels with different $f$ quantum number take place in 
this regime, as well as a repulsion among the magnetic sublevels with the same 
$f$ quantum number. 
This behaviour of the magnetic sublevels has important consequences on the 
atomic polarization, as pointed out by \citet{Bom80} and described in detail 
in LL04, and produce interesting effects, sometimes referred to as 
{\it level crossing effect} and {\it anti-level-crossing effect},
on the polarization signals produced by resonance scattering 
\citep[e.g.,][]{jtb02}.
In \S~\ref{sect:D2-vertical-field} we will show their effect on the
linear polarization of the Ba~{$\!$\sc ii} D$_{2}$ line.
If the magnetic field strength is further increased the 
so-called {\it complete Paschen-Back effect regime} is reached. 
In this regime the energy 
eigenvectors are of the form $|\alpha\, J\, I\, M_{J}\, M_{I}>$, and the 
splitting among the HFS magnetic sublevels is again linear with the magnetic 
field strength: the atom behaves in this regime as if it were devoid of HFS.
Going from the Zeeman effect regime to the complete Paschen-Back effect regime,
the magnetic field produces therefore an energy eigenvectors basis 
transformation.\\
\indent In the following we will work in the spherical statistical tensor
representation.
The conversion of the density matrix elements of equation~({\ref{eq:rho}})
into this representation is given by the relation (cf.~LL04)
\begin{eqnarray}
\rho_{\alpha\,J\,I}(i\,f,\,i^{\prime}\,f^{\prime}\,) = \sum_{FF^{\prime}}
C_{F}^{\,i}(\alpha\,J\,I,\,f)\,C_{F^{\prime}}^{\,i^{\prime}}
(\alpha\,J\,I,\,f^{\prime})\,
\rho_{\alpha\,J\,I}(F\,f,\,F^{\prime}\,f^{\prime}\,)=
\;\;\;\;\;\;\;\;\;\;\;\;\; & & \nonumber \\
=\sum_{FF^{\prime}}C_{F}^{\,i}(\alpha\,J\,I,\,f)\,
C_{F^{\prime}}^{\,i^{\prime}}(\alpha\,J\,I,\,f^{\prime})
\sum_{KQ}(-1)^{F-f}\sqrt{2K+1}
\Bigg( \begin{array}{ccc}
\!F & \!\!F^{\prime} & \!\!K \\
\!f & \!\!-f^{\prime} & \!\!-Q
\end{array} \!\!\Bigg)
\,^{\alpha\,J\,I}\rho^{K}_{Q}(F,F^{\prime}) \;\; . & &
\end{eqnarray}
The Statistical Equilibrium Equations (SEEs) and the radiative transfer 
coefficients for a multi-level atom with HFS, in the spherical statistical
tensor representation, written taking the quantization axis directed along
the magnetic field, can be found in \S~7.9 of LL04.  
Here we write only the expression for the emission coefficient in the 
transition between the upper level ($\alpha_u, J_u$) and the lower level 
($\alpha_\ell, J_\ell$)
\begin{eqnarray}
 & & \varepsilon_{j}(\nu,\mathbf{\Omega})=\frac{2h\nu^3}{c^2}
 \frac{h\nu}{4\pi}\mathcal{N} (2J_{u}+1)
 B(\alpha_{u}J_{u}I \to \alpha_{\ell} J_{\ell}I) \nonumber \\
 & & \times \sum_{KQK_{u}Q_{u}} \sqrt{3(2K+1)(2K_{u}+1)}
 \sum_{i_{u} F_{u} F_{u}^{\prime} F_{u}^{\prime\prime} i_{\ell}F_{\ell}
 F_{\ell}^{\prime}} \,\, \sum_{f_{u}f_{u}^{\prime}f_{\ell}qq^{\prime}}
 (-1)^{1+F_{u}^{\prime}-f_{u}+q^{\prime}} \nonumber \\
 & & \times C_{F_{\ell}}^{\,i_{\ell}}(\alpha_{\ell}J_{\ell}I,f_{\ell})\,
 C_{F_{\ell}^{\,\prime}}^{\,i_{\ell}}(\alpha_{\ell}J_{\ell}I,f_{\ell})\,
 C_{F_{u}}^{\,i_{u}}(\alpha_{u}J_{u}I,f_{u})\,
 C_{F_{u}^{\,\prime\prime}}^{\,i_{u}}(\alpha_{u}J_{u}I,f_{u}) \nonumber \\
 & & \times \sqrt{(2F_{\ell}+1)(2F_{\ell}^{\prime}+1)(2F_{u}+1)
 (2F_{u}^{\prime}+1)} \nonumber \\
 & & \times \Bigg( \begin{array}{ccc}
 F_{u} & F_{\ell} & 1 \\
 -f_{u} & f_{\ell} & -q
 \end{array} \Bigg)
 \Bigg( \begin{array}{ccc}
 F_{u}^{\,\prime} & F_{\ell}^{\,\prime} & 1 \\
 -f_{u}^{\,\prime} & f_{\ell} & -q^{\,\prime}
 \end{array} \Bigg)
 \Bigg( \begin{array}{ccc}
 1 & 1 & K \\
 q & -q^{\,\prime} & -Q
 \end{array} \Bigg) \nonumber \\
 & & \times \Bigg( \begin{array}{ccc}
 F_{u}^{\,\prime} & F_{u}^{\,\prime\prime} & K_{u} \\
 f_{u}^{\,\prime} & -f_{u} & -Q_{u}
 \end{array} \Bigg)
 \Bigg\{ \begin{array}{ccc}
 J_{u} & J_{\ell} & 1 \\
 F_{\ell} & F_{u} & I
 \end{array}\Bigg\}
 \Bigg\{\begin{array}{ccc}
 J_{u} & J_{\ell} & 1 \\
 F_{\ell}^{\,\prime} & F_{u}^{\,\prime} & I
 \end{array} \Bigg\} \nonumber \\
 & & \times {\rm{Re}} \Big[{\mathcal{T}}_{Q}^{K}(j,\mathbf{\Omega})\,
 ^{\alpha_{u}J_{u}I}\!\rho_{Q_{u}}^{K_{u}}(F_{u}^{\prime},F_{u}^{\prime\prime})
 \, \Phi(\nu_{\alpha_u J_u I i_u f_u, \alpha_{\ell} J_{\ell} I i_{\ell}
 f_{\ell}}-\nu) \Big] \; ,
\label{eq:epsilon}
\end{eqnarray}
where $j\!=\!0,1,2,3$, standing respectively for the Stokes parameters 
$I,Q,U$ and $V$, $\mathcal{N}$ is the number density of atoms, 
$B(\alpha_{u}J_{u}I \to \alpha_{\ell} J_{\ell}I)$ is the Einstein coefficient 
for stimulated emission, ${\mathcal{T}}_{Q}^{K}(j,\mathbf{\Omega})$ is a 
geometrical tensor (cf.~LL04), and $\Phi$ the profile of the line.\\
\indent It is important to note that the previous equations
are valid under the {\it{flat-spectrum approximation}}. 
For a multi-level atom with
HFS this approximation requires that the incident radiation
field should be flat (i.e.~independent of frequency) across a spectral interval
$\Delta\nu$ larger than the frequency intervals among the HFS levels (possibly 
split by the magnetic field), and larger than the inverse 
lifetimes of the same levels. 
This is a good approximation for the D$_1$ and D$_2$ lines of Ba~{$\!$\sc ii} 
if we restrict to magnetic fields smaller or of the order of 1~kG.

\subsection{{\it The Atomic Model}} 
\label{sect:atom}
\begin{table}[t!]
\label{tab:isotopes}
\begin{center}
{\footnotesize
\begin{tabular}{ccccccccc}
\tableline
\noalign{\smallskip}
Isotope & Abund. & $I$ & \multicolumn{2}{c}{Isotope Shifts (MHz)$^{1}$} & \multicolumn{4}{c}{HFS Constants (MHz)$^{2}$} \\
\noalign{\smallskip}
\cline{4-5}
\cline{6-9}
\noalign{\smallskip}
 & (\%)$^{\rm{a}}$ & & D$_1$ & D$_2$ & $^2S_{1/2}$ & $^2P_{1/2}$ & 
\multicolumn{2}{c}{$^2P_{3/2}$} \\
\noalign{\smallskip}
 & & & & & $\mathcal{A}$ & $\mathcal{A}$ & $\mathcal{A}$ & $\mathcal{B}$ \\
\tableline
\noalign{\smallskip}
$^{130}$Ba & 0.106 & 0 & 355.3$^{\rm{b}}$ & 372.3$^{\rm{b}}$ & & & \\
$^{132}$Ba & 0.101 & 0 & 278.9$^{\rm{b}}$ & 294.9$^{\rm{b}}$ & & & \\
$^{134}$Ba & 2.417 & 0 & 222.6$^{\rm{c}}$ & 234.6$^{\rm{c}}$ & & & \\
$^{135}$Ba & 6.592 & 3/2 & 348.6$^{\rm{b}}$ & 360.7$^{\rm{b}}$ & 3591.67$^{\rm{d}}$ & 664.6$^{\rm{e}}$ & 113.0$^{\rm{e}}$ & 59.0$^{\rm{e}}$ \\
$^{136}$Ba & 7.854 & 0 & 179.4$^{\rm{b}}$ & 186.9$^{\rm{b}}$ & & & \\
$^{137}$Ba & 11.232 & 3/2 & 271.1$^{\rm{b}}$ & 279.0$^{\rm{b}}$ & 4018.87$^{\rm{d}}$ & 743.7$^{\rm{e}}$ & 127.2$^{\rm{e}}$ & 92.5$^{\rm{e}}$ \\
$^{138}$Ba & 71.698 & 0 & \multicolumn{2}{c}{reference isot.} & & & \\
\noalign{\smallskip}
\tableline
\end{tabular}
}
\end{center}
\footnotesize{$^{1}$A positive I.S. means that the line is shifted to higher
frequencies with respect to the reference isotope.}\\
\footnotesize{$^{2}$The HFS constant $\mathcal{B}$ is defined according
to the convention of the American literature.}\\
\footnotesize{$^{\rm{a}}$NIST on-line database; $^{\rm{b}}$\citet{Wen84};
$^{\rm{c}}$\citet{Wen88}; $^{\rm{d}}$\citet{Bec81}; $^{\rm{e}}$\citet{Vil93}.}
\caption{\footnotesize{Isotopes considered in this work.}}
\end{table}
We adopt a three-level model of Ba~{$\!$\sc ii} consisting in the ground level
($6s\;^{2}S_{1/2}$), the upper level of the D$_{1}$ line ($6p\;^{2}P_{1/2}$) 
and the upper level of the D$_{2}$ line ($6p\;^{2}P_{3/2}$).
There are seven stable isotopes of barium, whose mass numbers and relative 
abundances are listed in Table~1. 
In this work we take into account the contributions coming from all 
the seven isotopes.\\
\indent The mass and volume differences between the nuclei of the various 
isotopes involve small but appreciable differences on the energies of the fine 
structure levels of different isotopes (isotopic effect).
We use the values of the isotopic shifts in the D$_{1}$ and D$_{2}$ 
lines listed in Table~1 to 
correct the energies of the $^{2}P_{1/2}$ and $^{2}P_{3/2}$ levels of the 
variuos isotopes. For the reference isotope
(138) we use the energy values given by \citet{Moore}.\\
\begin{figure}[!t]
\begin{center} 
\includegraphics[width=0.9\textwidth]{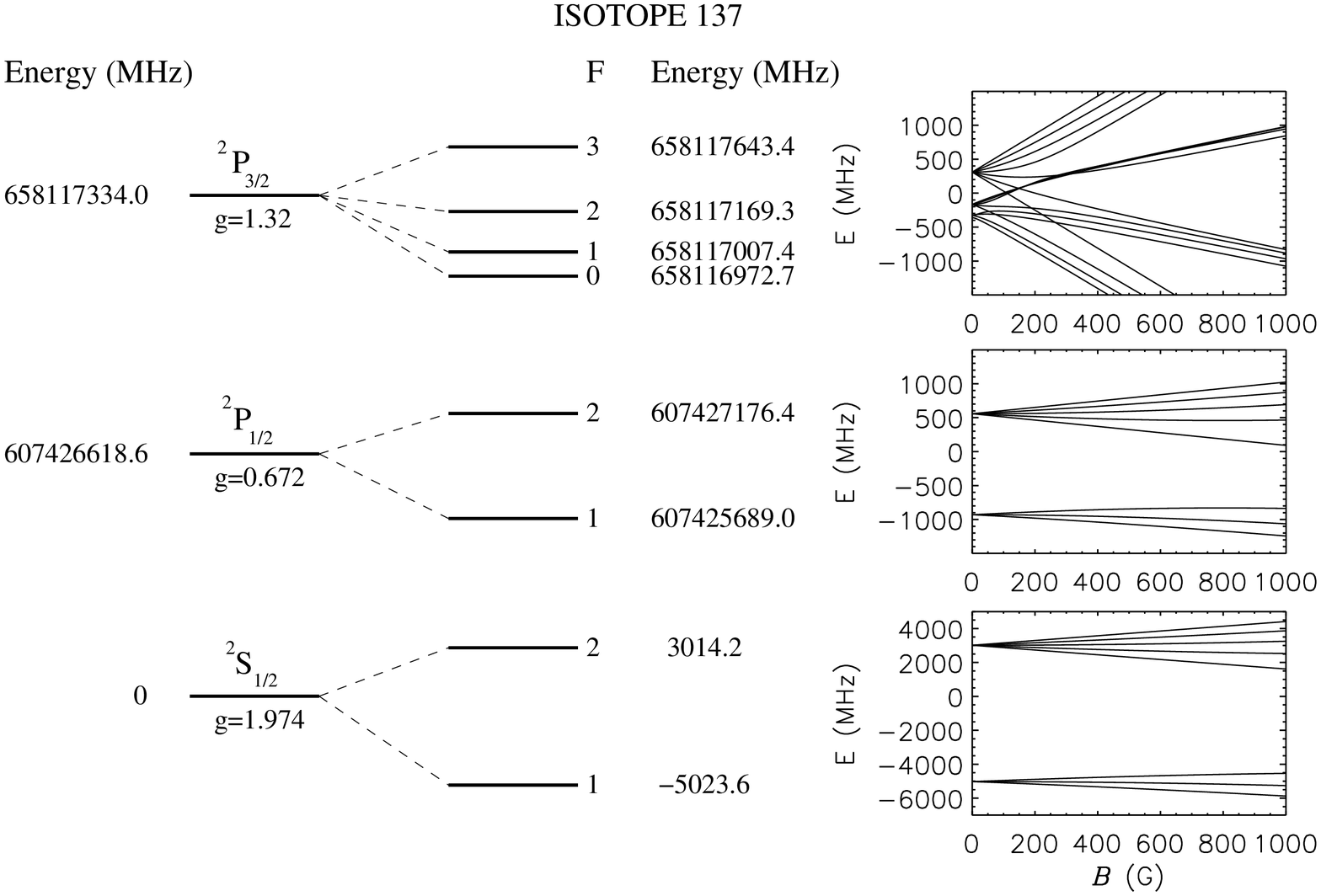}
\caption{{\footnotesize Energies, corrected for the isotopic effect, of the 
fine-structure and HFS levels of the isotope $^{137}$Ba~{\sc ii}. 
The right panels show the energies of 
the HFS magnetic sublevels as functions of the magnetic field strength 
(in each panel the zero of the energy scale is chosen at the energy of the 
corresponding fine structure $J$-level).
\label{fig:paschen}}}
\end{center}
\end{figure}
\indent Isotopes with even mass number have nuclear spin $I\!=\!0$, while 
those with odd mass number (135 and 137) have nuclear spin $I\!=\!3/2$. 
The odd isotopes show therefore HFS due to nuclear spin.
Introducing the total angular momentum, characterized by the quantum number
$F$, we observe (see Fig.~\ref{fig:paschen}) that the levels
$^{2}S_{1/2}$ and $^{2}P_{1/2}$ split into two HFS levels ($F\!=\!1,2$), 
while the level $^{2}P_{3/2}$ splits into four HFS 
levels ($F\!=\!0,1,2,3$).
It is possible to demonstrate that the HFS Hamiltonian can be 
expressed as an infinite series of electric and magnetic multipoles 
\citep[e.g.][]{Kop58}.
To calculate the energies of the various HFS levels we 
consider the first two terms of the series (magnetic dipole and electric 
quadrupole terms), and we use the HFS constants listed in 
Table~1.
This investigation will clearly show the importance of the HFS
effects for a correct modeling of the 
polarization produced by scattering processes in a stellar atmosphere.\\
\indent The Land\'e factors have been calculated theoretically, assuming L-S 
coupling for the Ba~{$\!$\sc ii} ion. The values obtained differ by less than 
1.3\% from the experimental ones \citep{Moore}, reported in 
Figure~\ref{fig:paschen}. 
This can be considered as a proof 
that the L-S coupling is quite a good approximation for the
Ba~{$\!$\sc ii} ion.\\
\indent The energies of the HFS magnetic sublevels of the isotope 
137, as functions of the magnetic field strength, are shown in 
Figure~\ref{fig:paschen}.
We can see that in the range between 0 and 1000~G the splitting 
of the various magnetic sublevels originating from the ground level 
$^{2}S_{1/2}$ is linear with the magnetic field strength (Zeeman effect 
regime). A similar behaviour is shown by the magnetic sublevels originating 
from the level $^{2}P_{1/2}$ for magnetic fields smaller than about 600~G. 
For stronger magnetic fields the linearity of the splitting appears to be 
slowly lost, which indicates that the incomplete Paschen-Back effect regime is 
reached.
The splitting observed among the magnetic sublevels originating from the level 
$^{2}P_{3/2}$ shows instead that a complete transition from the Zeeman effect 
regime to the complete Paschen-Back effect regime takes place for magnetic 
fields ranging from 0 to 1000~G.
As described in \S~\ref{sect:equations}, 
in the intermediate incomplete Paschen-Back effect regime
several level crossings among HFS magnetic sublevels can be observed, 
as well as a repulsion among the sublevels with the same 
$f$ quantum number.

\subsection{{\it The Incident Radiation Field}}
\label{sect:radiation}
As already stated in \S~\ref{sect:formulation}, we consider an optically 
thin plane-parallel slab, composed of Ba~{$\!$\sc ii} ions, located at 
approximately 1000~km above the $\tau_{5000}\!=\!1$ photospheric level, 
and we assume that the slab is illuminated from below 
(hence, anisotropically) by the photospheric continuum radiation. 
Under these hypothesis the atomic polarization can be calculated solving 
directly the SEEs for the given continuum radiation field coming from 
the photosphere.\\
\begin{table}[t!]
\label{tab:transitions}
\smallskip
\begin{center}
{\footnotesize
\begin{tabular}{ccccc}
\tableline
\noalign{\smallskip}
Line & $\lambda$~(\AA) & $A\,(s^{-1})$ & $\bar{n}_{\nu}$ & $w_{\nu}$ \\
\noalign{\smallskip}
\tableline
\noalign{\smallskip}
 D$_1$ & 4934.09 & 0.955$\times 10^8$ & 0.323$\times 10^{-2}$ & 0.159 \\
 D$_2$ & 4554.03 & 1.17$\times 10^8$ & 0.225$\times 10^{-2}$ & 0.176 \\
\noalign{\smallskip}
\tableline
\end{tabular}
\caption{\footnotesize{Wavelength (in air), and Einstein coefficient of the 
transitions considered; mean number of photons and anisotropy factor of the 
photospheric continuum at the wavelength of the transitions, 1000~km above 
the $\tau_{5000}=1$ level.}}
}
\end{center}
\end{table}
\indent Let us take a reference system with the {\it{z}} axis directed along the
local vertical, and let us describe the continuum radiation field incident on 
the slab by means of the tensor 
\begin{equation}
J^{K}_{Q}(\nu)=\int\frac{{\rm d}\Omega}{4\pi}\sum_{i=0}^{3}
{\mathcal{T}}_{Q}^{K}(i,\mathbf{\Omega}) S_i(\nu,\mathbf{\Omega}) \:\: ,
\end{equation}
where $S_{i}=I,Q,U,V$.
Assuming that the incident radiation field is unpolarized and has cylindrical
symmetry around the local vertical, it is easy to verify that the only non
vanishing components are
\begin{equation}
J^{0}_{0}(\nu)\! \! =\! \! \oint\frac{{\rm{d}}\Omega}{4\pi}I(\nu,\mu) 
\;\;\;\;\; {\rm{and}} \;\;\;\;\; 
J^{2}_{0}(\nu)\! \! =\! \! \oint\frac{{\rm{d}}\Omega}{4\pi}\Big(\frac{1}
{2\sqrt{2}} (3\mu^2-1)I(\nu,\mu) \Big) \;\; .  
\label{eq:JKQ}
\end{equation}
Note that $J^{0}_{0}$ is just the mean intensity of the incident radiation
(averaged over all directions), while $J^{2}_{0}$ gives a measure of 
the anisotropy of the radiation field\footnote{In particular $J_{0}^{2}$ 
quantifies the unbalance between vertical and horizontal illuminations.}.
Instead of $J^{0}_{0}$ and $J^{2}_{0}$, we can use the quantities 
$\bar{n}(\nu)$, the mean number of photons, and $w_{\nu}$, the so called 
anisotropy factor. The new quantities are related to the previous radiation 
field tensor components through the relations
\begin{equation}
\bar{n}(\nu)\! =\! \frac{c^{2}}{2h\nu^{3}}J^{0}_{0}(\nu) 
\;\;\;\;\; {\rm and} \;\;\;\;\;
w_{\nu}\! =\! \sqrt{2}\,\frac{J^{2}_{0}(\nu)}{J^{0}_{0}(\nu)} \;\; .
\end{equation}
To calculate the values of $\bar{n}(\nu)$ and $w_{\nu}$ of the photospheric 
continuum at the height of 1000~km above the visible solar ``surface'', at the 
frequencies of the D$_{1}$ and D$_{2}$ lines of Ba~{$\!$\sc ii}, we follow 
\S~12.3 of LL04.
The values of the specific intensity of the radiation coming from the solar 
disk center and of the limb-darkening coefficients are taken from 
\citet{Allen}. The values obtained for $\bar{n}_{\nu}$ and $w_{\nu}$ are listed 
in Table~2.
At this point the SEEs can be solved numerically.
Their expression becomes simpler if we rewrite them in the reference system
with the quantization axis directed along the local vertical direction, as
in this case only two components of the radiation field tensor are non zero
($J_{0}^{0}$ and $J_{0}^{2}$).
It can be demonstrated that all the radiative rates are invariant under a
rotation of the reference system so that only 
the magnetic kernel has to be modified with respect to the expression 
given in equation~(7.66) of LL04.

\section{THE POLARIZATION OF THE ATOMIC LEVELS} 
\label{sect:polarization}
\begin{figure}[!t] 
\begin{center}
\includegraphics[angle=270,width=0.9\textwidth]{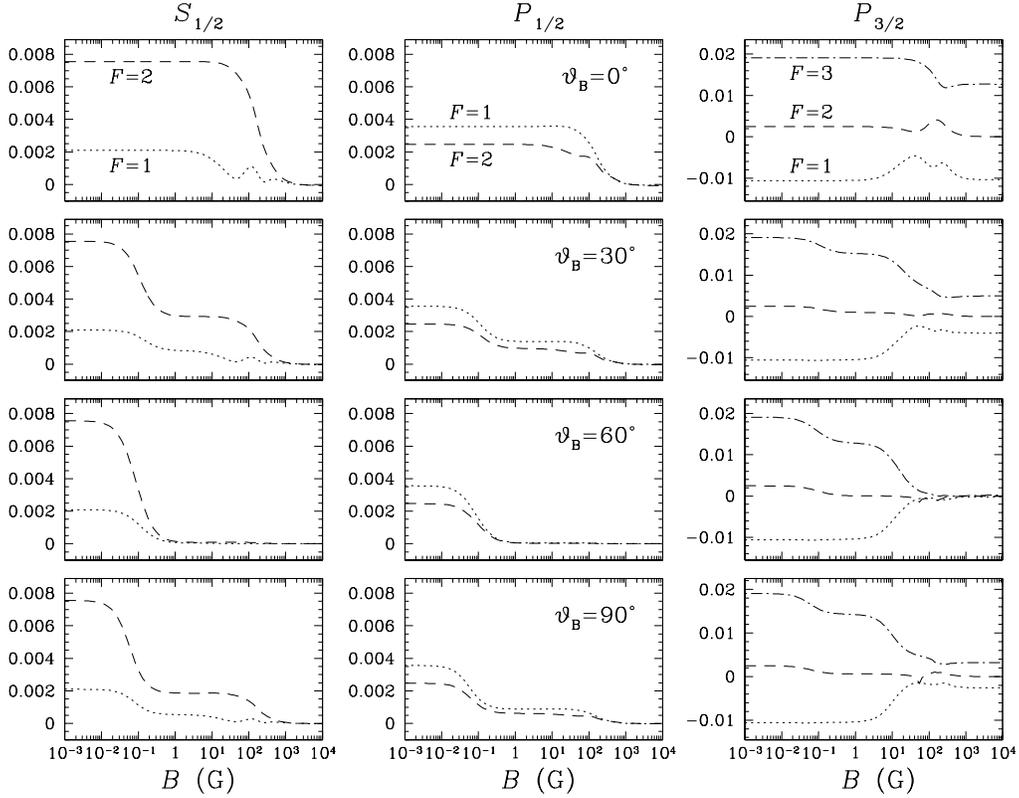}
\caption{{\footnotesize Fractional atomic alignment, 
$\sigma^{2}_{0}=\rho^{2}_{0}(F,F) / \rho^{0}_{0}(F,F)$, of the various 
HFS levels of the isotope 137, calculated in the local 
vertical reference system, as functions of the magnetic field vector, for 
various inclination angles. Top row $\theta_{B}\!=\!0^{\circ}$, middle rows 
$\theta_{B}\!=\!30^{\circ}$ and $\theta_{B}\!=\!60^{\circ}$, bottom row 
$\theta_{B}\!=\!90^{\circ}$. The angle $\theta_B$ is defined in 
Figure~\ref{fig:geometry}.}}
\label{fig:sigma20}
\end{center}
\end{figure}

We solve numerically the SEEs 
(which implies, for each isotope with HFS of our 
model atom, the solution of a linear system of 384 equations in the unknowns 
$\rho^K_Q(F,F^{\,\prime}\,)$) for magnetic field strengths between 0 and 
1000~G, and for various inclinations of the magnetic field with respect to the 
local vertical.
We recall that the $\rho^0_0(F,F\,)$ elements quantify the populations of the 
various $F$-levels, the $\rho^2_Q$ elements ({\it{alignment}} components) 
contribute to the linear polarization of the scattered radiation, while 
the $\rho^{1}_{Q}$ elements ({\it{orientation}} components) contribute to the 
circular polarization of the scattered radiation.
As shown in LL04, an anisotropic, unpolarized, 
flat spectrum radiation field generally induces only 
alignment in the atomic system, while orientation can be
originated by the so called alignment-to-orientation conversion mechanism.
Note that all the levels of the isotopes 135 and 137, because of the HFS, 
can carry alignment while, for all the other isotopes, 
only the level $^2P_{3/2}$, the upper level of the D$_{2}$ line, can carry it.\\
\indent Let us consider isotope 137.  In complete analogy with 
the case of Na~{$\!$\sc i}, having a single isotope with $I\!=\!3/2$ 
\citep[see Trujillo Bueno et al.~2002 and][and the discussion therein]{Cas02},
only the level $^2P_{3/2}$ is polarized directly 
via the anisotropic illumination. The ground level becomes polarized because of 
a transfer of polarization via spontaneous emission in the D$_{2}$ line, while 
the level $^2P_{1/2}$ becomes polarized via radiative absorption
({\it{repopulation pumping}}) in the D$_{1}$ line. This explains the fact that
the upper and lower levels of the D$_{1}$ line are equally sensitive to the
magnetic field strength, independently of its inclination 
(see Fig.~\ref{fig:sigma20}).\\
\indent It is well known that a magnetic field is able to modify the atomic 
polarization, and therefore the polarization of the scattered radiation
(Hanle effect\footnote{\label{note:Hanle}In this work by Hanle 
effect we mean any modification of the atomic polarization which is due to 
the action of a magnetic field. Note, however, that in the literature it is 
often meant by Hanle effect only the relaxation of coherences (defined in the 
magnetic field reference system) having $Q\ne0$.
Within this second meaning, it is often stated that there is no Hanle effect 
in the presence of a vertical magnetic field. Note that according to our 
definition this statement is true only if we are dealing with an 
isolated level (i.e.~if we neglect the quantum interferences among the magnetic 
sublevels originating from different hyperfine (or fine) structure levels).
We prefer to adopt the former more general definition because, as we are 
dealing with a quite complex atomic system with HFS, and as we are 
investigating the role of magnetic fields with  
complex configurations (random-azimuth and microturbulent), it becomes quite 
difficult to understand which effects due to the magnetic field can be 
considered as `Hanle effect' according to the latter definition.}). 
The behaviour of the various spherical statistical tensors, 
written in the local vertical reference system, as functions of the magnetic 
field inclination and strength, is qualitatively equal to the case of 
Na~{$\!$\sc i}, investigated by \citet{jtb02}.
We point out that the first decrease of the atomic polarization of 
the D$_1$ levels, which takes place only for inclined fields 
(see Fig.~\ref{fig:sigma20}), is due to the so called lower-level Hanle 
effect\footnote{This definition has been formulated within the latter 
definition of the Hanle effect given in the footnote~\ref{note:Hanle}.}.
As can be demonstrated (see LL04 for details), the spherical 
statistical tensors $\rho^K_Q(F,F)$ (that describe population imbalances and 
quantum interferences among magnetic sublevels originating from the same 
$F$-level, and that mainly affect the polarization of the scattered radiation 
at the line center) are significantly modified by a magnetic field when the 
Zeeman splitting is of the same order of magnitude as the inverse lifetime of 
the level. That is, as a rough estimation, when the magnetic field ranges 
between (see LL04) 
\begin{equation}
\label{eq:formulita}
0.1\,B_{\rm c} \le B \le 10 \, B_{\rm c} \;\; ,
\end{equation}
with
\begin{equation}
B_{\rm c}\approx\frac{1.137\times10^{-7}}{t_{\rm life}\,g_{\rm L}} \;\; ,
\end{equation}
where $t_{\rm life}$ (in seconds) is the radiative lifetime of the lower 
or upper level of the line transition under consideration, $g_{\rm L}$ 
is its Land\'e factor and $B_{\rm c}$ is the critical magnetic field intensity
in G\footnote{The previous expression of $B_{\rm c}$ is exact only for an 
isolated level (see footnote\ref{note:Hanle}).}. 
Since the relevant atomic level here
is the ground level, it is important to note that its radiative lifetime is 
$t_{\rm life} \approx 1/(B_{\ell u}\, J_{0}^{0})$. 
As seen in Figure~\ref{fig:sigma20} this 
first decrease takes place for magnetic fields of the order of $10^{-1}$G, 
consistently with our simplifying assumption that the pumping radiation field 
is the continuum radiation tabulated by \citet{Allen}. If, on the contrary, one 
takes into account the line profile (resulting in a smaller value of 
$J_{0}^{0}$) the decrease will take place for smaller magnetic fields 
\citep[cf.~the Na~{$\!$\sc i} results of][]{jtb02}. 
Note that this decrease of the atomic polarization in the ground level 
has a feedback even on the D$_2$ upper level.
Concerning the second sudden decrease of the atomic polarization of the 
D$_1$ levels, for $B$ larger than 100~G, we recall that this is due to the
inhibition of the repopulation pumping mechanism discussed by \citet{jtb02} 
and by \citet{Cas02}, which sets in when the electronic and nuclear angular 
momenta, $J$ and $I$, of the $^{2}P_{3/2}$ level are decoupled, 
(i.e.~when this level, the only one that can carry alignment even in the 
absence of HFS, enters the complete Paschen-Back effect regime). 
Note that for the case
of sodium this sudden decrease occurs for $B$ larger than 10~G simply because 
the complete Paschen-Back effect regime is reached for weaker magnetic fields 
in sodium than in barium.\\ 
\indent As expected, because of the symmetry of the problem,
for a vertical magnetic field only the components
with $Q\!=\!0$ are non zero. For different orientations of the magnetic field 
we have in general contributions coming from all the density matrix elements;
in particular it is possible to demonstrate that the
components with $Q\!=\!0$ are independent of the magnetic field azimuth, 
the components with $Q\!=\!1$ change sign under an azimuth rotation of 
$180^{\circ}$, the components with $Q\!=\!2$ change sign under an azimuth 
rotation of $90^{\circ}$, and so on.\\
\indent The expressions of the emission coefficients (eq.~\ref{eq:epsilon}), 
as well as the expressions of all the other radiative transfer coefficients 
given in LL04 hold in the magnetic reference system.
Therefore we have to transform the spherical statistical tensors, obtained
solving the SEEs written in the local vertical 
reference system, into the magnetic field reference system.
Indicating with $[\rho^{K}_{Q}(F,F^{\,\prime}\,)]_{B}$ the spherical statistical
tensor components in the magnetic field reference system, and with
$[\rho^{K}_{Q}(F,F^{\,\prime}\,)]_{V}$ the spherical statistical tensor 
components in the local vertical reference system, we have
\begin{equation}
[\rho^{K}_{Q}(F,F^{\,\prime}\,)]_{B}=
\sum_{Q^{\prime}}\,[\rho^{K}_{Q^{\prime}}(F,F^{\,\prime}\,)]_{V}\,
D^{K}_{Q^{\prime}Q}(R)^{\ast} \;\; ,
\end{equation}
where $D^{K}_{Q^{\prime}Q}(R)$ is the rotation matrix calculated for the
rotation $R$ which carries the vertical reference system into the magnetic
reference system (referring to Fig.~{\ref{fig:geometry}}, we have 
$R\!=\!(\chi_B,\theta_B,0)$), and where the apex 
``$\,\ast\,$'' indicates the complex conjugate.
We observe that after the rotation, because of the symmetry of the problem,
the spherical statistical tensors in the magnetic reference
system do not depend on the azimuth but only on the inclination of the
magnetic field with respect to the local vertical reference system.

\section{THE POLARIZATION OF THE EMERGENT SPECTRAL LINE RADIATION} 
\label{sect:emergent}
For the case of a tangential observation in a plane-parallel atmosphere, it 
can be shown that, under the approximation of a weakly polarizing
atmosphere ($\varepsilon_{I}\gg\varepsilon_{Q},\varepsilon_{U},
\varepsilon_{V}$; $\eta_{I}\gg\eta_{Q},\eta_{U},\eta_{V}$),
the emergent fractional polarization is given by 
\citep[e.g.,][]{jtb03}
\begin{equation}
\label{eq:dichroism}
\frac{X(\nu,\mathbf{\Omega})}{I(\nu,\mathbf{\Omega})} \approx
\frac{\varepsilon_{X}(\nu,\mathbf{\Omega})}
{\varepsilon_{I}(\nu,\mathbf{\Omega})}-
\frac{\eta_{X}(\nu,\mathbf{\Omega})}
{\eta_{I}(\nu,\mathbf{\Omega})} \;\;\;\;\;\; {\rm{with}}\; X=Q,U,V \;\; .
\end{equation}
We point out that the first term of equation~(\ref{eq:dichroism}) is the 
contribution to 
the emergent fractional polarization due to selective emission processes, while 
the second one is caused by dichroism (selective absorption of polarization 
components).
As shown in Figure~\ref{fig:sigma20}, for the D$_{2}$ line the contribution due 
to dichroism is much smaller than that due to selective emission. 
For this reason, from now on we will describe the polarization properties of 
the radiation emergent from the slab using the relation
\begin{equation}
\frac{X(\nu,\mathbf{\Omega})}{I(\nu,\mathbf{\Omega})}=
\frac{\varepsilon_{X}(\nu,\mathbf{\Omega})}
{\varepsilon_{I}(\nu,\mathbf{\Omega})} \;\;  .
\label{eq:fract-polar1}
\end{equation}
We recall that the polarization properties of the
emergent radiation will always be described assuming the reference direction 
for positive $Q$ parallel to the slab.\\
\indent It is important to remember that the expressions for the emission and 
absorption coefficients given in LL04 take into account only the line processes.
For this reason, in order to be able to reproduce qualitatively the 
observed profiles, we need to add the contribution coming from the continuum.
Assuming that the continuum is not polarized and constant across
the line, we have
\begin{equation}
\frac{X(\nu,\mathbf{\Omega})}{I(\nu,\mathbf{\Omega})}=
\frac{\varepsilon_{X}^{\, l}(\nu,\mathbf{\Omega})}
{\varepsilon_{I}^{\, l}(\nu,\mathbf{\Omega})+\varepsilon^{\,\rm c}_{I}} \;\;  ,
\label{eq:fract-polar2}
\end{equation}
where $\varepsilon^{\,\rm c}_{I}$ is the continuum contribution to the 
intensity of the emergent radiation, and where the apex ``{ \it l} '' is to 
recall that the corresponding quantity refers only to the line processes.
In Appendix~\ref{app:dichroism} we show and briefly discuss some results
obtained by applying equation~(\ref{eq:dichroism}).\\
\indent The Ba~{$\!$ \sc ii} D-lines that we are investigating are strong lines:
according to theoretical models of the solar atmosphere the wings of these 
lines originate in the photosphere, while the line-cores in the high 
photosphere-low chromosphere.
The optically thin slab model illuminated by the photospheric continuum 
that we are considering in this paper is therefore just a zero-order 
approximation. 
Nevertheless, it allows us to take into account 
in a very rigorous way the atomic physics involved in the problem, and to 
understand its essential role on the magnetic sensitivity of the polarization 
profiles of these lines, avoiding complications coming from radiation transfer 
effects. This is the first step of our investigation, in a forthcoming paper 
we will propose more realistic models, where radiation transfer effects
will be taken into account.\\
\indent Once the SEEs in the vertical reference system 
have been solved numerically, and 
the spherical statistical tensors have been rotated to the magnetic
field reference system, we can calculate the 
emission coefficients for $90^{\circ}$ scattering 
($\theta \!= \! 90^{\circ}$ and $\chi \!= \! 0^{\circ}$ in 
Fig.~\ref{fig:geometry}) 
by means of equation~(\ref{eq:epsilon}), and the polarization of the 
scattered radiation through equation~(\ref{eq:fract-polar2}). In the 
following subsections we present our results for the Ba~{$\!$\sc ii} D$_2$ and 
D$_1$ lines, for various magnetic field configurations.

\subsection{{\it D$_2$ Line -- No Magnetic Field Case:\\
Origin of the Three Peaks Structure and Choice
of Parameters Values}}
\label{sect:D2-zero-field}
\begin{figure}[!t] 
\begin{center}
\plottwo{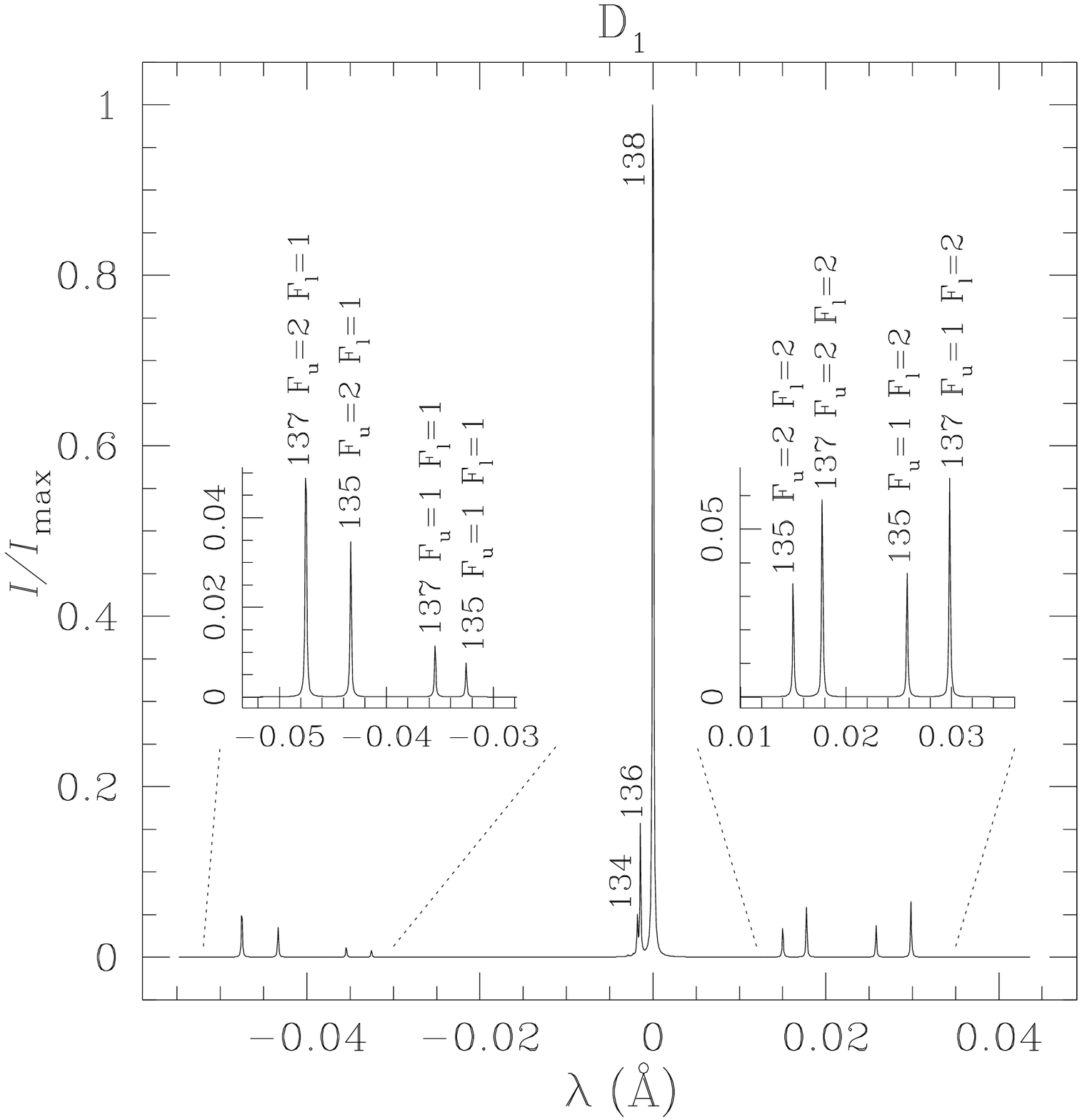}{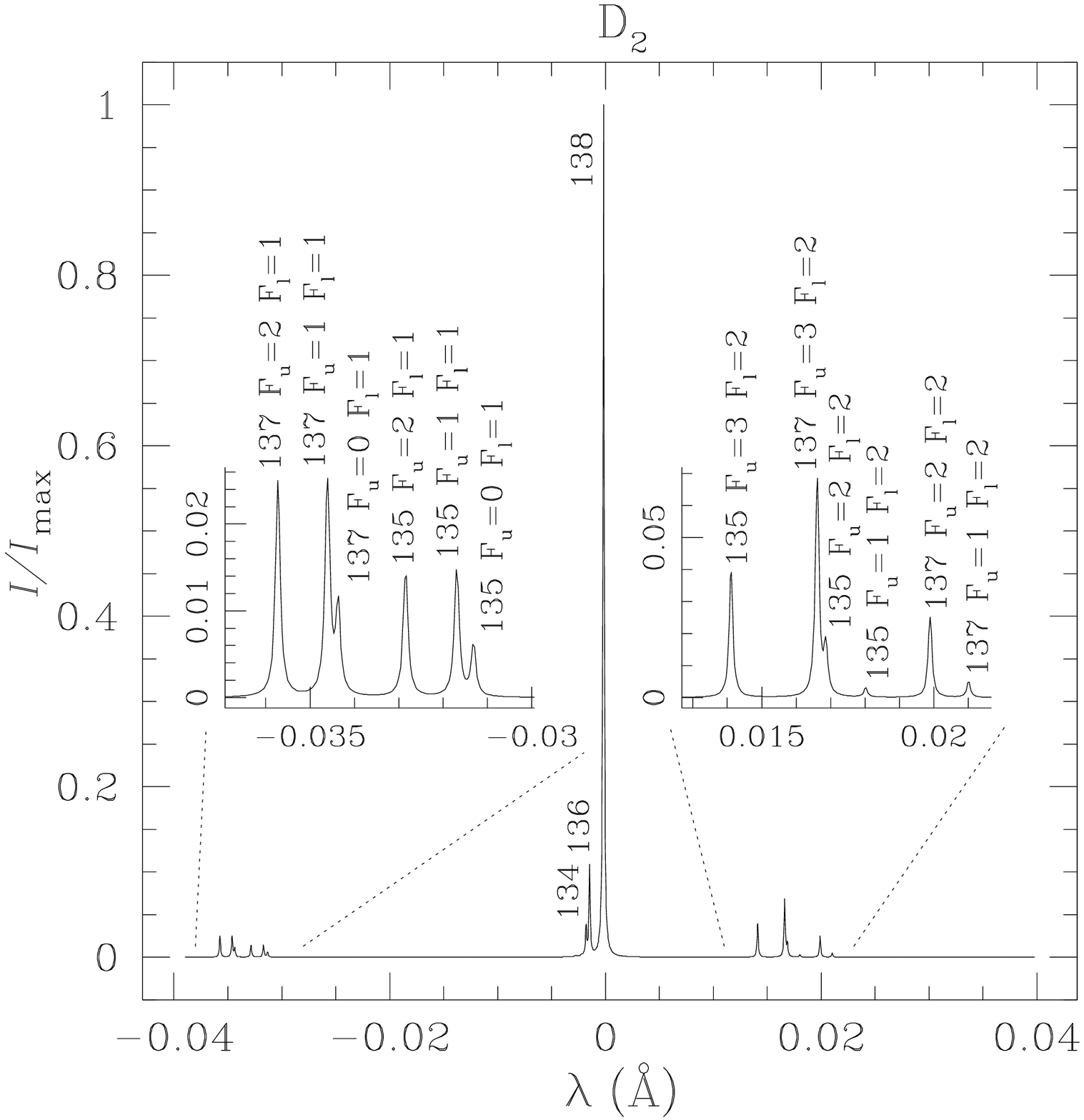}
\caption{{\footnotesize Laboratory positions of the various 
HFS components of the D$_{1}$ (left panel) and D$_{2}$ (right panel) lines 
resulting from the seven isotopes considered. The horizontal axis gives the 
wavelength (in \AA) measured from 4934.075~{\AA} for the D$_{1}$ line,
and from 4554.029~{\AA} for the D$_{2}$ line. 
These are the wavelengths (in air) of the isotope 138
D$_1$ and D$_2$ lines, respectively.}}
\label{fig:hfscomp}
\end{center}
\end{figure}

\begin{figure}[!t] 
\begin{center}
\includegraphics[angle=270, width=0.95\textwidth]{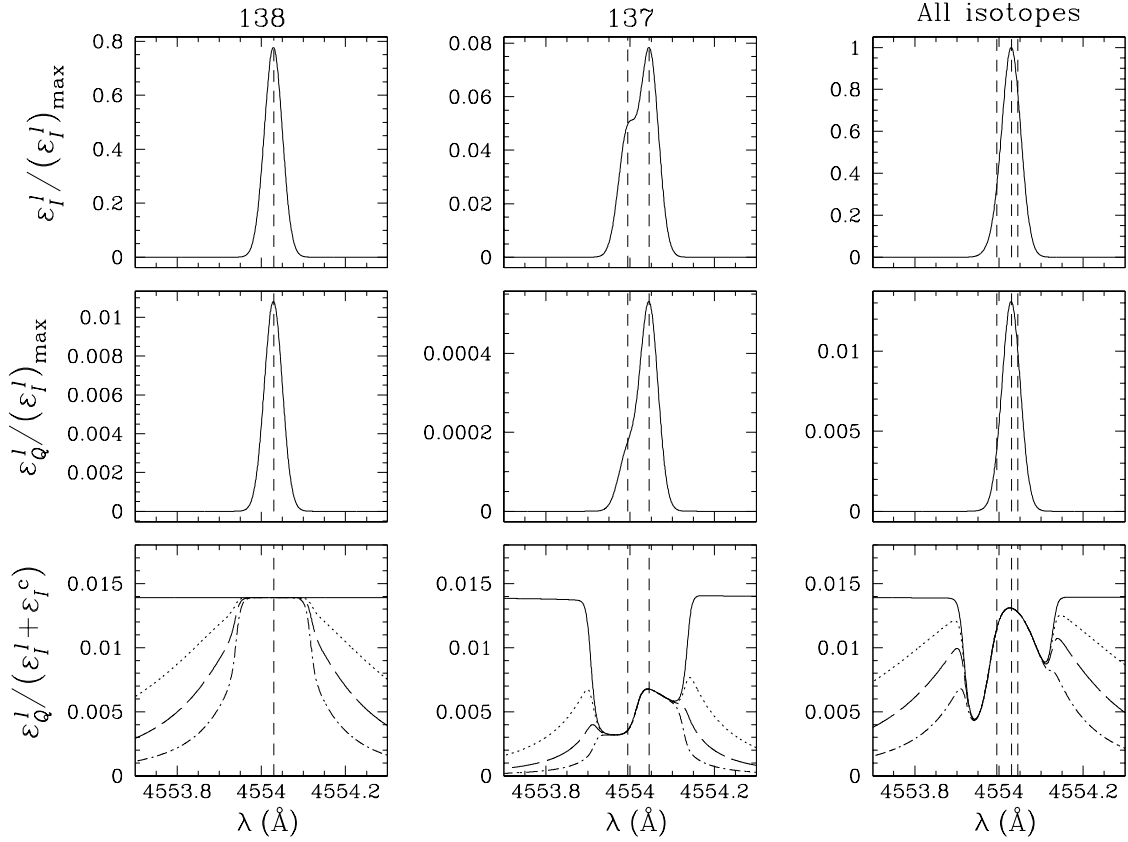}
\caption{{\footnotesize Theoretical profiles of the emission coefficients 
$\varepsilon_{I}^{\,l}(\lambda)$ and $\varepsilon_{Q}^{\,l}(\lambda)$,
and of the ratio 
$\varepsilon_{Q}^{\,l}(\lambda)/(\varepsilon_{I}^{\,l}(\lambda)+
\varepsilon_{I}^{\, \rm c})$, of the Ba~{\sc ii} D$_2$ line, for the isotope 
138, for the isotope 137 and for all the seven isotopes together, in the 
absence of a magnetic field. The emission profiles are 
normalized to the maximum value of the intensity emission profile, 
calculated taking into account the contribution of all the seven isotopes, 
$(\varepsilon_{I}^{l})_{\rm{max}}$.  
The vertical dashed lines show the positions of the three groups 
of transitions (see text). 
The last row shows the 
$\varepsilon_{Q}^{\,l}(\lambda)/(\varepsilon_{I}^{\,l}(\lambda)+
\varepsilon_{I}^{\, \rm c})$ profiles, without continuum (solid), 
with a continuum $\varepsilon_{I}^{\, \rm c}/(\varepsilon_{I}^{l})_{\rm{max}}$ 
of $10^{-5}$ (dot), of  
$3\times 10^{-5}$ (long dash), and of
$9\times 10^{-5}$ (dash-dot).
\label{fig:D2-zero-field}}}
\end{center}
\end{figure}

As a first step, we observe the laboratory positions of the various 
HFS components of the D$_{2}$ line. The three isotopes without HFS contribute 
to the D$_2$ line with just one component each, those which have HFS 
contribute to this line with six components each, which may overlap.
As seen in Figure~\ref{fig:hfscomp} (right panel), it is possible to divide the 
various HFS components into three groups. The central group, at about 
4554.03{\AA}, is composed by the five components (three visible in the figure) 
due to the five isotopes without HFS (note that the main contribution comes 
from the isotope 138 because of its high abundance, while isotopes 130 and 132 
bring a negligible contribution, not visible in the figure). 
The other components, due to the isotopes with HFS, fall at different 
wavelengths but can be gathered into two groups
because of the large splitting of the ground level into the two $F\!=\!1$ and 
$F\!=\!2$ HFS levels (see Fig.~\ref{fig:paschen}). In particular, the group at 
about 4553.995{\AA} is composed by the HFS components, of both isotopes 135 
and 137, associated to the transitions towards the lower $F\!=1\!$ HFS level, 
while the group at about 4554.045{\AA} is composed by the HFS components
associated to the transitions to the lower $F\!=\!2$ HFS level.
As we will see in more detail below, and as already
pointed out by \citet{Ste97b}, the origin of the three peaks structure of 
the D$_{2}$ line lies in this splitting of the various components into these 
three groups. Similar considerations
could be done about the position and relative strength of the various
components of the D$_{1}$ line, with the difference that now the upper 
level just has two HFS levels (instead of four).\\
\indent Let us begin our analysis of the D$_{2}$ profile by considering only 
the isotope 138 (without HFS), and let us assume a Doppler width of 30~m{\AA}. 
For this isotope, as shown in the left panels of 
Figure~\ref{fig:D2-zero-field}, 
the ratio $\varepsilon_{Q}^{\, l}(\lambda)/\varepsilon_{I}^{\, l}(\lambda)$ 
(often referred to as fractional polarization)
is constant and different from zero\footnote{See Appendix~\ref{app:two-level} 
for an analytical proof of this result.}. 
Adding the contribution of the continuum, the same ratio remains unchanged in 
the line-core (where $\varepsilon_{I}^{\, l}\gg\varepsilon_{I}^{\, \rm c}$), 
while it goes to zero at the wavelengths corresponding to the wings of the 
intensity profile.
The same considerations can be done for all the other isotopes without HFS.
Let us consider now the isotope 137 (with HFS). For this
isotope, assuming the same Doppler width, the profiles of the line emission 
coefficients $\varepsilon_{I}^{\, l}$ and 
$\varepsilon_{Q}^{\, l}$ show two peaks at the wavelength positions
of the two groups of HFS transitions (see Fig.~\ref{fig:D2-zero-field}).
The ratio $\varepsilon_{Q}^{\, l}(\lambda)/\varepsilon_{I}^{\, l}(\lambda)$ is
no longer constant but decreases showing two broad minima at the wavelength 
positions corresponding to the wings of the $\varepsilon_{I}^{\, l}$ and 
$\varepsilon_{Q}^{\, l}$ profiles, and assumes the same value as the isotopes
without HFS moving away from the line-core\footnote{See 
Appendix~\ref{app:two-level} for an analytical proof of this result.}. 
This profile clearly shows the depolarizing effect of the HFS.
The role of the continuum is the same as observed for the isotopes without
HFS. The same arguments hold for the isotope 135.\\
\indent This investigation on the isotopes 138 and 137 shows that the
central peak of the observed $Q/I$ profile is due to the isotopes without 
HFS, while the two secondary peaks are due to the isotopes with HFS. 
The position and amplitude of these secondary peaks appear to be strongly 
dependent on the background continuum emissivity $\varepsilon_{I}^{\,\rm c}$,
a physical quantity that, given the exploratory character of this paper, we will
just parametrize in order to reproduce at best the observed profile.\\
\indent We consider now all the isotopes together and we adjust the
Doppler width, the anisotropy factor and the continuum
intensity in order to obtain the best fit to the $Q/I$ profile observed by
\citet{Ste97a}, still assuming that no magnetic field is present.
Changing the Doppler width we can modify the separation between the two 
minima of the profile
$\varepsilon_Q^{\, l}(\lambda)/\varepsilon_I^{\, l}(\lambda)$. 
To obtain the same separation as the observed profile we need a value of 
about 30~m{\rm{\AA}}. This value seems
to be very reasonable as it can be obtained assuming a temperature of
about 6000~K and a microturbulent velocity of about 1.8~km/s, values which are
in good agreement with those given by semi-empirical chromospheric models 
at the height of about 1000~km.\\
\indent Modifying the anisotropy factor we simply scale the ratio
$\varepsilon_Q^{\, l}(\lambda)/\varepsilon_I^{\, l}(\lambda)$.
Radiative transfer effects are disregarded in our model and, since the
Ba~{$\!$\sc ii} D$_{2}$ line is a strong line, we can expect that the 
calculated value for the anisotropy factor taking into account only the 
photospheric continuum (see \S~\ref{sect:radiation}),
is probably overestimated.
\begin{figure}[!t]
\begin{center}
\includegraphics[width=0.70\textwidth]{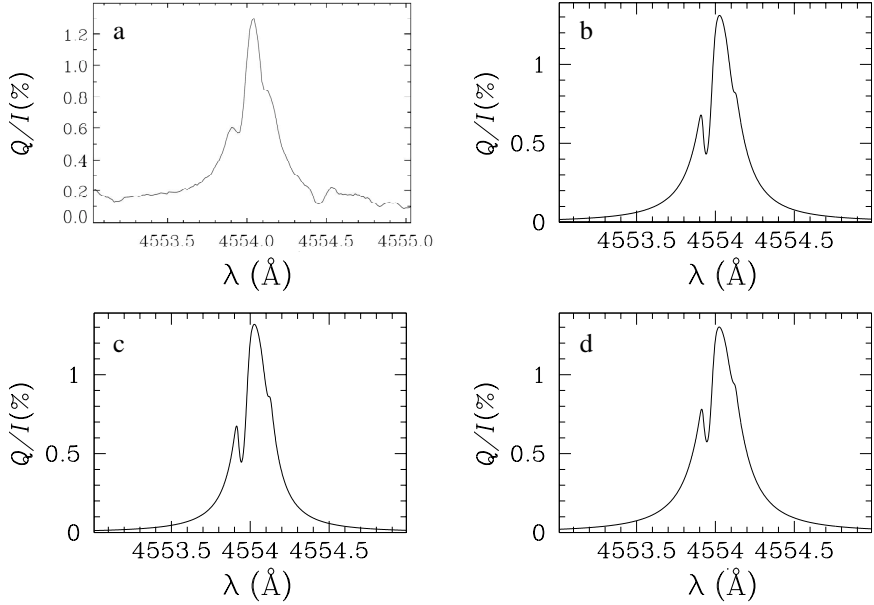}
\caption{{\footnotesize {\it Panel} a: observed $Q/I$ profile of the 
Ba~{\sc ii} D$_{2}$ line \citep{Ste97a}.
{\it Panel} b: theoretical $Q/I$ profile
obtained in the absence of a magnetic field, choosing a Doppler width
($\Delta \lambda_D$) of 30~m{\rm{\AA}}, an anisotropy factor ($w$) of 0.037,
and a continuum $\varepsilon_{I}^{\,\rm c}/(\varepsilon_{I}^{l})_{\rm{max}}$ of
$9 \times 10^{-5}$.
{\it Panel} c: theoretical $Q/I$ profile obtained in the presence of a
microturbulent magnetic field of 5~G, choosing the following values of the free
parameters: $\Delta \lambda_D$=30~m{\rm{\AA}}, $w$=0.052,
$\varepsilon_{I}^{\,\rm c}/(\varepsilon_{I}^{l})_{\rm{max}}$=$1.77 \times
10^{-4}$.
{\it Panel} d: theoretical $Q/I$ profile obtained in the presence of a
vertical magnetic field of 40~G, choosing the following values of the free
parameters: $\Delta \lambda_D$=30~m{\rm{\AA}}, $w$=0.037,
$\varepsilon_{I}^{\,\rm c}/(\varepsilon_{I}^{l})_{\rm{max}}$=
$7.5\times 10^{-5}$.}}
\label{fig:bestfit}
\end{center}
\end{figure}
We find indeed that the anisotropy factor has to be decreased to the value of 
0.037 (approximately $1/5$ of the value 0.176 mentioned in 
\S~\ref{sect:radiation}) to obtain the observed value of the ratio 
$Q/I$ at the wavelength position of the central peak.\\
\indent In this paper we are not taking into account the continuum processes 
in a rigorous way. As mentioned above, we describe the effect of the 
continuum simply through the parameter $\varepsilon^{\,\rm c}_{I}$. 
As shown in Figure~\ref{fig:D2-zero-field}, the continuum modifies the 
wings of the profile. Taking as a reference
the ``red'' secondary peak of the observed profile, we find that the best fit
is obtained assuming a value for 
$\varepsilon_{I}^{\,\rm c}/(\varepsilon_{I}^{l})_{\rm{max}}$ of
$9\times 10^{-5}$.
With these values of the parameters, applying equation~(\ref{eq:fract-polar2}), 
we get the profile shown in the panel b of Figure~\ref{fig:bestfit} which 
reproduces quite well the 
observed profile, and which is very similar to a theoretical profile already
obtained by \citet{Ste97b}.\\
\indent The main aim of this work is to investigate the magnetic 
sensitivity of the 
linear and circular polarization of the D-lines of Ba~{$\!$\sc ii}, and not
merely to reproduce as better as possible the observed profiles of these lines.
For this reason, we prefer to perform our investigation by sticking 
to the values of the parameters found for the simpler, unmagnetized case. 
Obviously there is 
no reason to think that the best agreement with the observed profile has to be 
found in the absence of a magnetic field. It is likely enough that the best 
agreement could be obtained in the presence of a deterministic or
microturbulent magnetic field, using different values of the parameters. 
For example, in panels c and d of Figure~\ref{fig:bestfit} we show the best
theoretical profiles that we have obtained in the presence of a microturbulent
magnetic field of 5~G and a vertical magnetic field of 40~G, choosing different
values of the free parameters (see caption to Fig.~{\ref{fig:bestfit}}).

\subsection{{\it D$_2$ Line -- The Influence of a Magnetic Field on the 
Emergent Polarization}}
\label{sect:magnetic}
Depending on its strength, and on its direction with respect to the local 
vertical and to the direction of the scattered radiation, a magnetic field 
will differently modify the linear and circular polarization of the line 
through the Zeeman and the Hanle effects\footnote{Hereafter, regardless of the 
particular regime (Zeeman effect regime, 
incomplete or complete Paschen-Back effect regime), any polarization signal 
that originates from the splitting among the magnetic sublevels will be 
referred to as Zeeman effect.}. 
It is well known that the Zeeman effect produces 
in general elliptical polarization, which degenerates into linear polarization 
if the magnetic field lies on the plane perpendicular to the line-of-sight 
(LOS), and into circular polarization if the magnetic field lies along the 
LOS. The Zeeman effect dominates the polarization of the scattered radiation 
if the splitting among the magnetic sublevels is of the same order of magnitude 
or larger than the Doppler width of the line. This criterion gives a critical 
value of the magnetic field strength for the Ba~{$\!$\sc ii} D$_2$ line of 
about 3000~G. However, if the magnetic field 
is not too weak, and if there are no other mechanisms that dominate the 
polarization, it is possible to identify Zeeman effect signatures on the 
fractional polarization profiles even for intensities much smaller than the 
critical value. As we will see below in 
Figures~\ref{fig:D2-vertical-field} and \ref{fig:D2-vertical-Zeeman}, for 
the line under investigation magnetic fields of about 50~G
are enough in order the transverse Zeeman effect to produce 
appreciable modifications of the linear polarization signal.\\ 
\indent On the other hand, as described in \S~\ref{sect:polarization}, 
a magnetic field is able to modify the atomic polarization, and therefore the 
polarization of the scattered radiation.
Depending on the configuration of the magnetic field, and on the geometry of 
the scattering event, different signatures of the Hanle effect can be
produced on the polarization profiles. 
As the observed scattering polarization in the Ba~{$\!$\sc ii} 
D$_2$ line is dominated by the atomic polarization of the upper level,
recalling that for the $^2P_{3/2}$ level
$t_{\rm life} \approx 1/A_{u\ell} \approx 10^{-8}$~s and $g_{\rm L}=1.33$,
applying equation~(\ref{eq:formulita}), we find that the line
is expected to be sensitive to the Hanle effect for magnetic field 
strengths ranging approximately between 1~G and 100~G (values which are 
smaller than the ones needed for the transverse Zeeman effect to be 
appreciable).

\subsection{{\it D$_2$ Line -- Vertical Magnetic Field}}
\label{sect:D2-vertical-field}
\begin{figure}[!t]
\plottwo{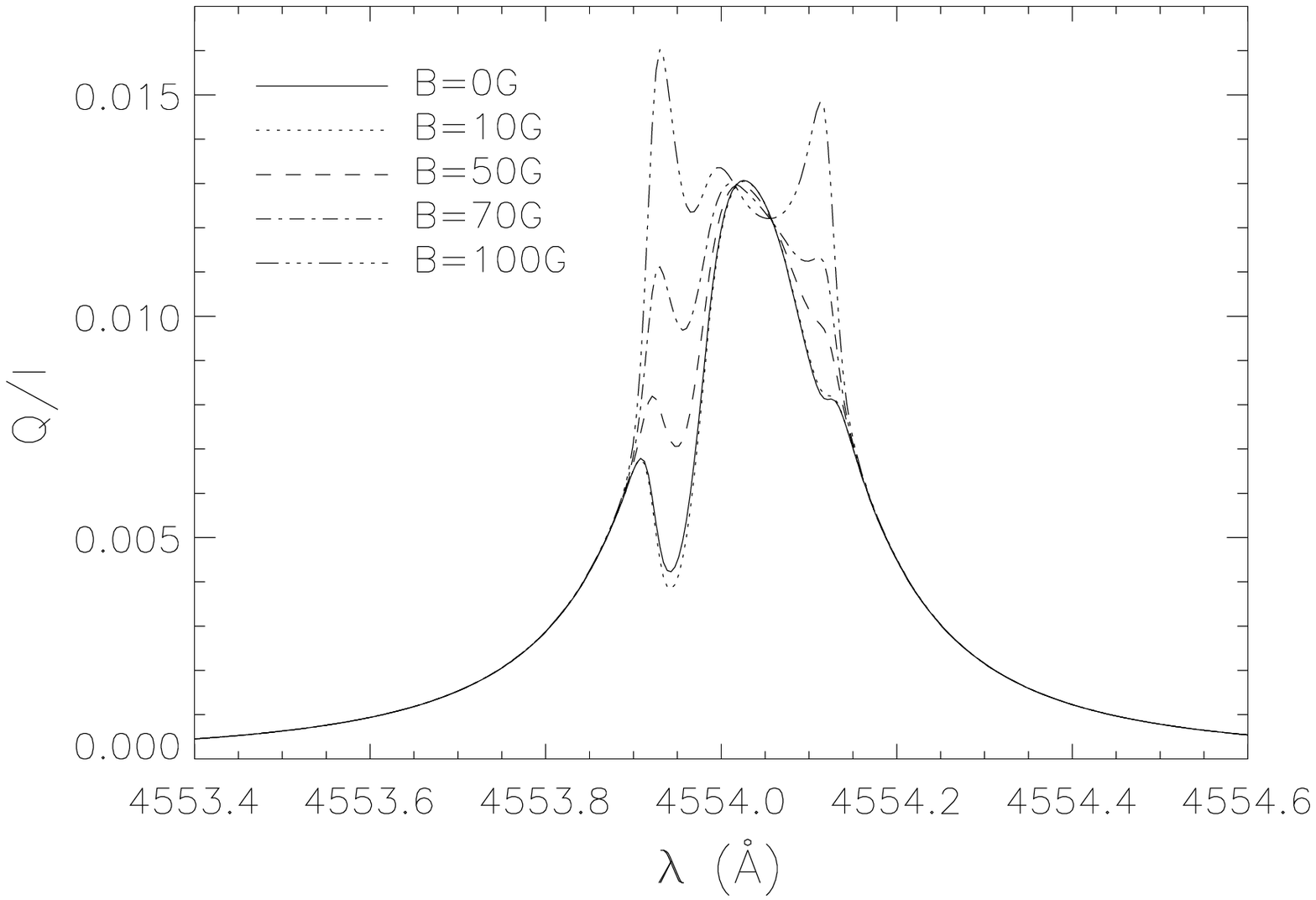}{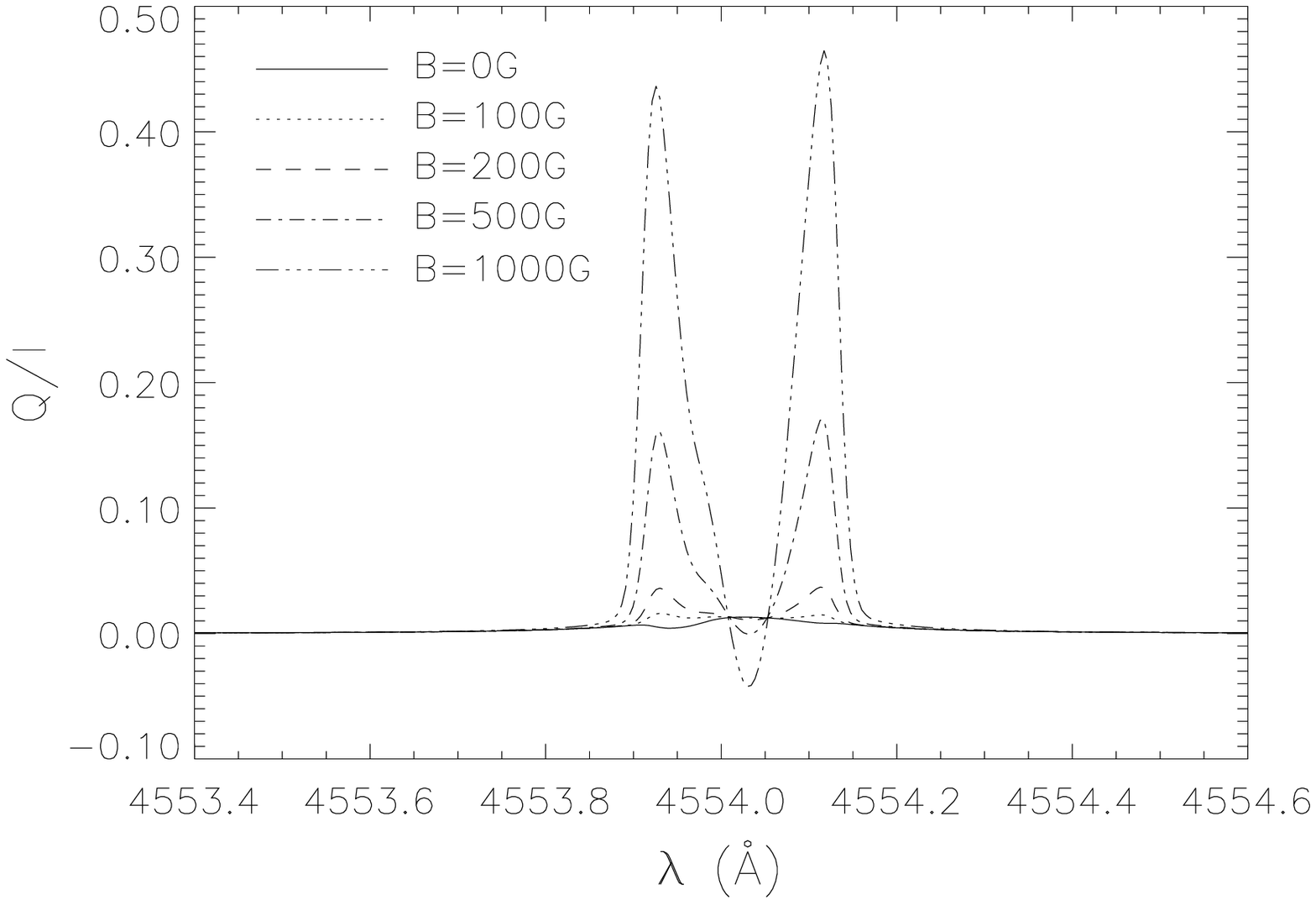}
\caption{{\footnotesize Theoretical $Q/I$ profile of the Ba~{\sc ii} D$_2$ line 
in the presence of a vertical magnetic field ($\theta_B \!=\! 0^{\circ}$). 
In the left panel the magnetic field varies between 0 and 100~G while
in the right panel it varies between 0 and 1000~G. Note the difference in the 
scale of the two figures.}}
\label{fig:D2-vertical-field}
\end{figure}

In this section we consider the effect of a vertical magnetic field on the 
theoretical $Q/I$ profile\footnote{Hereafter by `theoretical $Q/I$ profile' 
we will always refer to the profile obtained by applying 
equation~(\ref{eq:fract-polar2}).} of Figure~\ref{fig:bestfit} (panel b). 
The results are shown in Figure~\ref{fig:D2-vertical-field}.
The first interesting feature is the enhancement of the linear polarization 
at the wavelength positions of the two dips between the line-core peak and 
the two secondary peaks, for magnetic fields relatively 
weak (less than 50~G), for which the influence of the transverse 
Zeeman effect is negligible. 
In order to understand which physical mechanism is at the origin of this and
other features shown by these $Q/I$ profiles, in the various ranges of magnetic 
field intensity, we try to distinguish which polarization properties of the 
emergent radiation are due to the atomic polarization effects and which 
ones are due to the Zeeman effect. 
To this aim, we can obtain 
interesting information by plotting the profiles obtained through 
equation~(\ref{eq:fract-polar2}) according to two different strategies: 
\begin{itemize}
\item[{\sc a)}] taking the nominal values for the 
spherical statistical tensors, but setting $B\!=\!0$ when calculating the energy
eigenvalues and eigenvectors,
\item[{\sc b)}] setting equal to zero all the 
spherical statistical tensors, except $\rho^{0}_{0}$, but taking properly into 
account the influence of $B$ on the energy eigenvalues and eigenvectors. 
\end{itemize}
\begin{figure}[!t]
\includegraphics[width=0.95\textwidth]{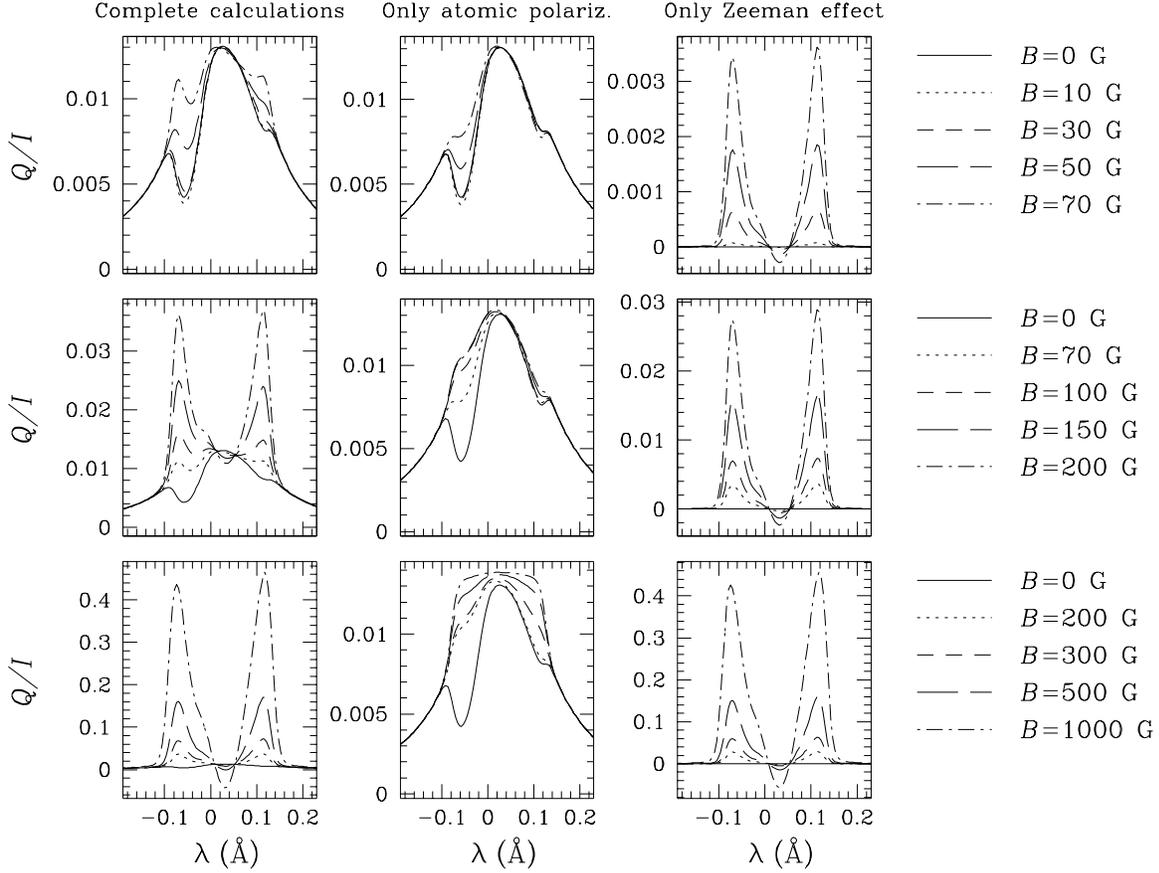}
\caption{{\footnotesize {\it Left column}: theoretical $Q/I$ profiles of the 
Ba~{\sc ii} D$_{2}$ line obtained in the presence of a vertical magnetic field. 
{\it Middle column}: theoretical $Q/I$ profiles obtained neglecting the Zeeman 
effect (only atomic polarization effects). 
{\it Right column}: theoretical $Q/I$ profiles obtained neglecting atomic 
polarization effects (only Zeeman effect). The zero of the wavelength scales 
is taken at 4554~\AA.}}
\label{fig:D2-vertical-Zeeman}
\end{figure}
In the former case (Fig.~\ref{fig:D2-vertical-Zeeman}, middle column) we are 
taking into account only the effects due to atomic polarization, neglecting 
the Zeeman effect, while in the latter case 
(Fig.~\ref{fig:D2-vertical-Zeeman}, right 
column) we are taking into account only this second effect, within the 
framework of the Paschen-Back effect theory.\\
\indent From Figure~{\ref{fig:D2-vertical-Zeeman}}, it is clear that the 
above-mentioned enhancement of the linear polarization at the wavelength 
positions of the dips, in the presence of a weak vertical magnetic field, 
is not due to the Zeeman effect, but to the Hanle effect acting only on 
the isotopes with HFS (see \S~\ref{sect:D2-zero-field}). 
We note in fact that the central peak, which is due to the isotopes without HFS 
is not sensitive, in this range, to the magnetic field.
Actually this particular behaviour can be explained in terms of two different 
mechanisms: the anti-level-crossing effect (briefly introduced in 
\S~\ref{sect:equations}), and the change of coupling scheme 
of the atomic system\footnote{According to our definition, these effects should 
be better considered as particular cases of the Hanle effect.}.
We note first that if the magnetic field lies along the symmetry 
axis of the radiation field, only the statistical tensors $\rho_0^0$ and 
$\rho_0^2$ are different from zero (see also \S~\ref{sect:radiation}). 
As the incomplete Paschen-Back effect regime is reached, the HFS magnetic 
sublevels with the same $f$ quantum number separate from each other
(see \S~\ref{sect:equations}), as a consequence the terms
$\rho_0^2(F,F^{\prime})$ (which 
quantifies the corresponding quantum interferences) decrease, and 
this causes an increase of the polarization of the scattered radiation 
(anti-level-crossing-effect, see LL04 for details).
On the other hand, as already stated in \S~\ref{sect:equations}, going 
from the Zeeman effect regime to the complete Paschen-Back effect regime, 
the magnetic field produces an energy eigenvectors basis transformation. 
This transformation implies a changing of the coupling scheme of the atomic 
system, which affects the polarization state of the atomic 
system\footnote{The complex 
mechanism of inhibition of atomic polarization transfer discussed in 
\citet{Cas02} is a particular consequence of this coupling scheme 
transformation.}.
Both these mechanisms begin to play an appreciable role as the upper level of 
the D$_2$ enters the incomplete Paschen-Back effect regime.
The order of magnitude of the magnetic field strength needed to reach 
this regime can be estimated from the relation
$0.1 \le \Delta\lambda_{B}/\Delta\lambda_{{\rm{hfs}}} \approx 1$, where
$\Delta\lambda_{B}$ is the splitting induced by the magnetic field, and
$\Delta\lambda_{{\rm{hfs}}}$
is the wavelength separation between the HFS $F$-levels.
Applying this relation to the Ba~{$\!$\sc ii} D$_2$ line, we find that this 
effect is expected to take place for magnetic fields larger than 10~G, as it is
observed in Figure~\ref{fig:D2-vertical-Zeeman}.
The possibility of an enhancement of the scattering polarization in the
presence of a vertical magnetic field, through this kind of mechanisms,
was already pointed out by \citet{jtb02} for the case of the
Na~{$\!$\sc i} D$_2$ line.\\
\indent Increasing the magnetic field strength, the transverse 
Zeeman effect eventually 
becomes appreciable and, besides the previous effect, we see an increase of 
the polarization at the wavelength position of the two peaks on the wings of 
the profile (see Fig.~\ref{fig:D2-vertical-field} and 
Fig.~\ref{fig:D2-vertical-Zeeman}).
Going to magnetic fields of about 200~G or stronger we enter the transverse 
Zeeman effect regime, and the linear polarization profile takes the typical 
symmetrical shape\footnote{Note that here we are plotting the ratio
$\varepsilon_{Q}(\lambda)/\varepsilon_{I}(\lambda)$ and not just
$\varepsilon_{Q}(\lambda)$.} (see Fig.~\ref{fig:D2-vertical-field} 
and Fig.~\ref{fig:D2-vertical-Zeeman}).

\subsection{{\it D$_2$ Line -- Horizontal Magnetic Field, Perpendicular to 
the Line of Sight}}
\label{sect:D2-horizontal-field}
\begin{figure}[!t]
\plottwo{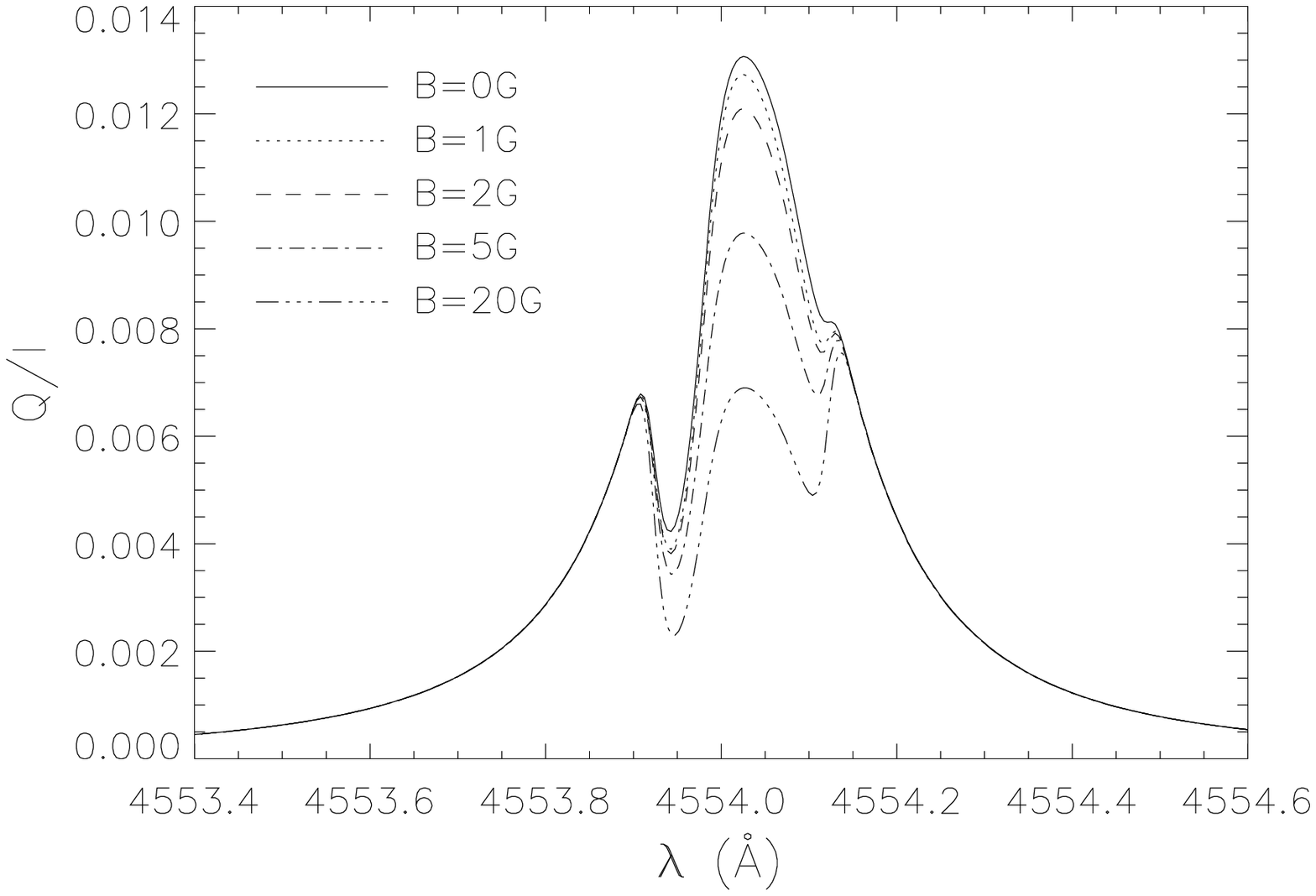}{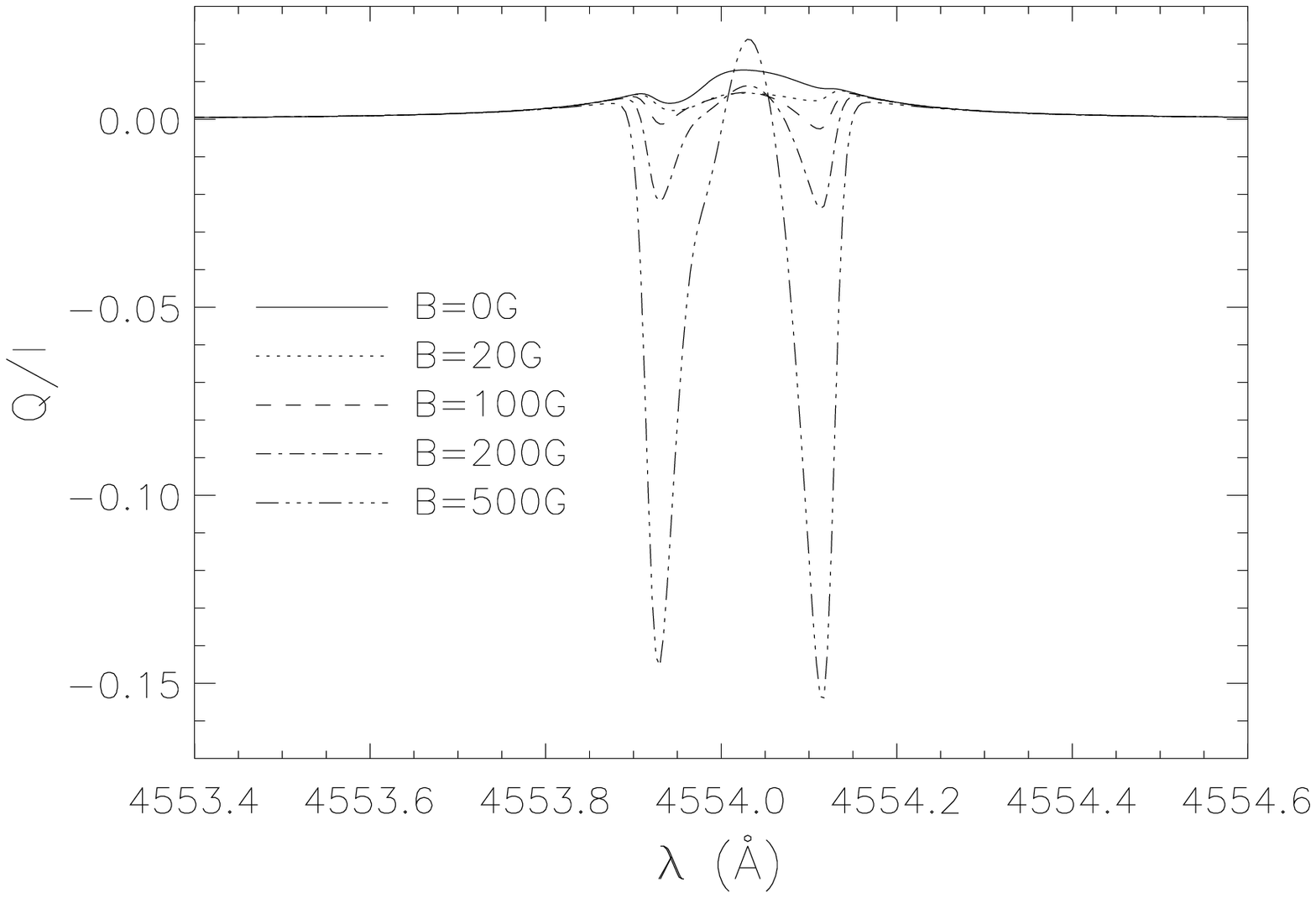}
\caption{{\footnotesize Theoretical $Q/I$ profiles of the Ba~{\sc ii} D$_2$ line
in the presence of a horizontal magnetic field, perpendicular to the line of 
sight ($\theta_B \!=\! 90^{\circ}, \, 
\chi_B \!=\! \pm 90^{\circ}$)}}
\label{fig:D2-horizontal-field}
\end{figure}
In the presence of a weak horizontal magnetic field, perpendicular
to the line of sight, as the field increases we observe
a decrease of the linear polarization at the wavelength position of the
central peak, due to the Hanle effect.
However, consistently with the fact that the Hanle effect has to vanish in
the far wings of the line, the two peaks on the wings remain almost unaffected,
as can be seen in the left panel of Figure~\ref{fig:D2-horizontal-field}.
Similarly to what happens in the presence of a vertical magnetic field,
going to intensities of about 50~G or stronger we enter the
transverse Zeeman effect regime, and the $Q/I$ profile takes the well known
shape shown in the right panel of Figure~\ref{fig:D2-horizontal-field}.

\subsection{{\it D$_2$ Line -- Horizontal Magnetic Field, Directed Along the 
Line of Sight}}
\label{sect:D2-longitudinal-field}
\begin{figure}[!t]
\plottwo{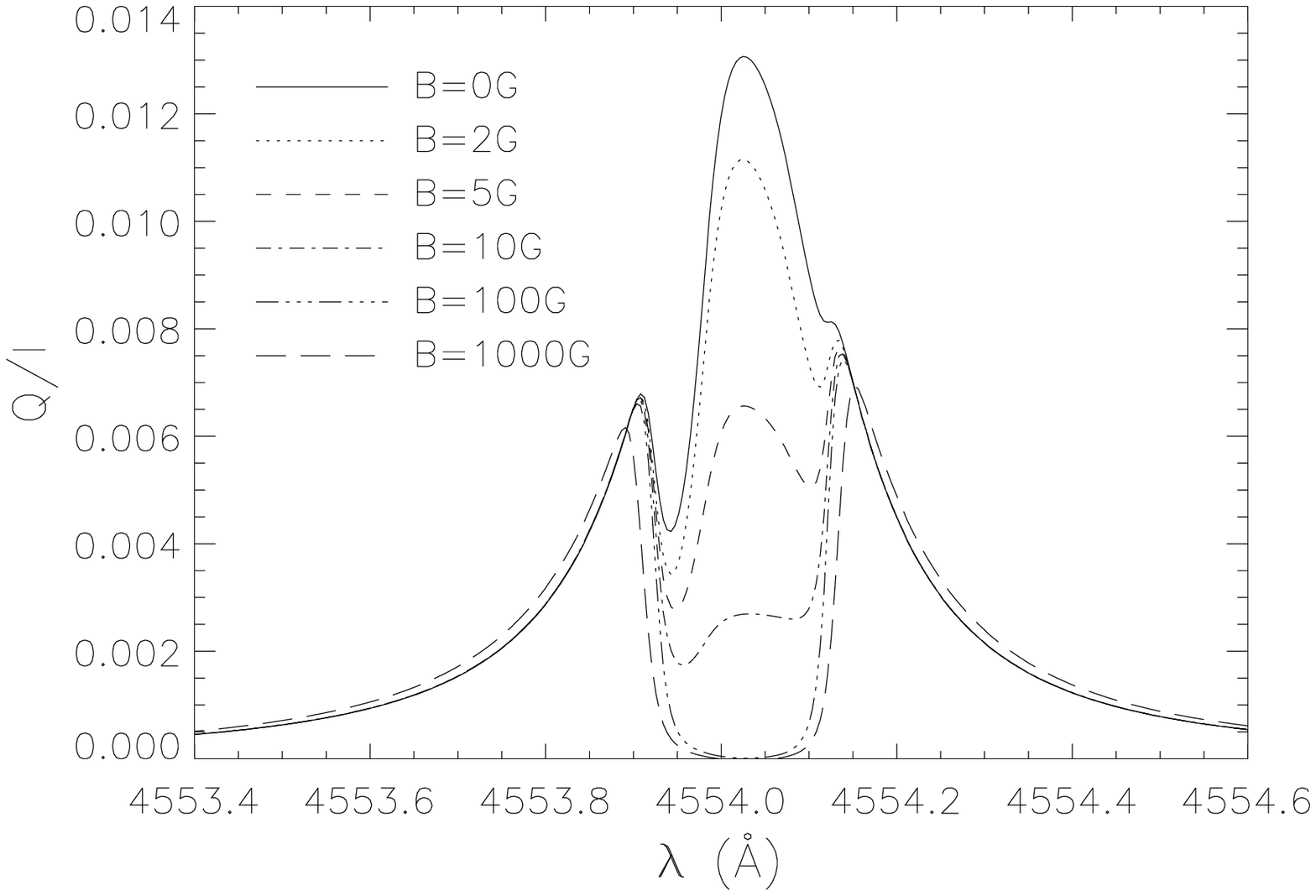}{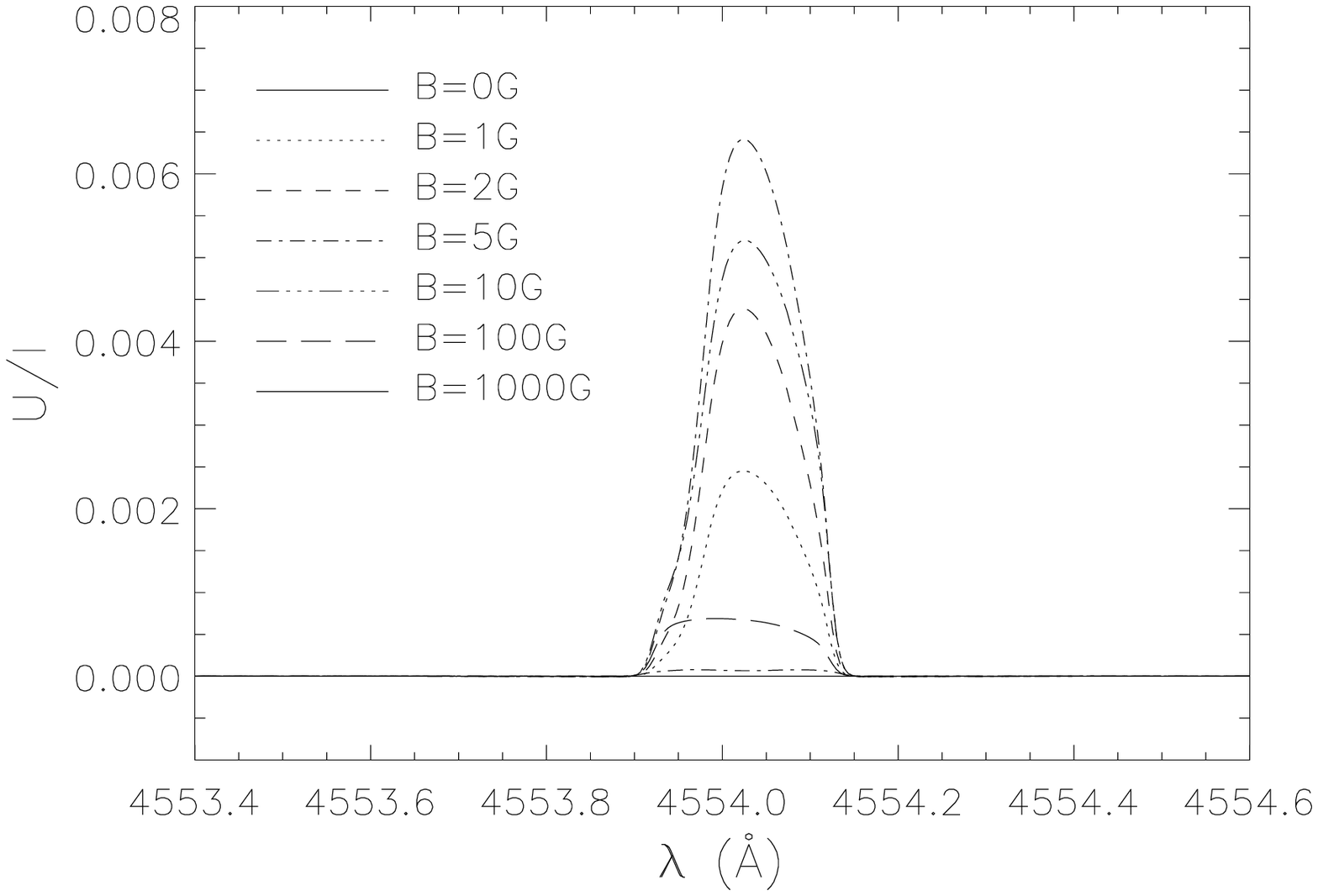}
\caption{{\footnotesize Theoretical $Q/I$ (left panel) and $U/I$
(right panel) profiles of the Ba~{\sc ii} D$_2$ line in 
the presence of a longitudinal magnetic field ($\theta_B \!=\! 90^{\circ}, \, 
\chi_B \!=\! 0^{\circ}$). For $\chi_B \! = \! 180^{\circ}$ the $U/I$ profile 
would be the same except for an overall sign switch.}}
\label{fig:D2-longitudinal-field-QU}
\end{figure}
\begin{figure}[!t]
\plottwo{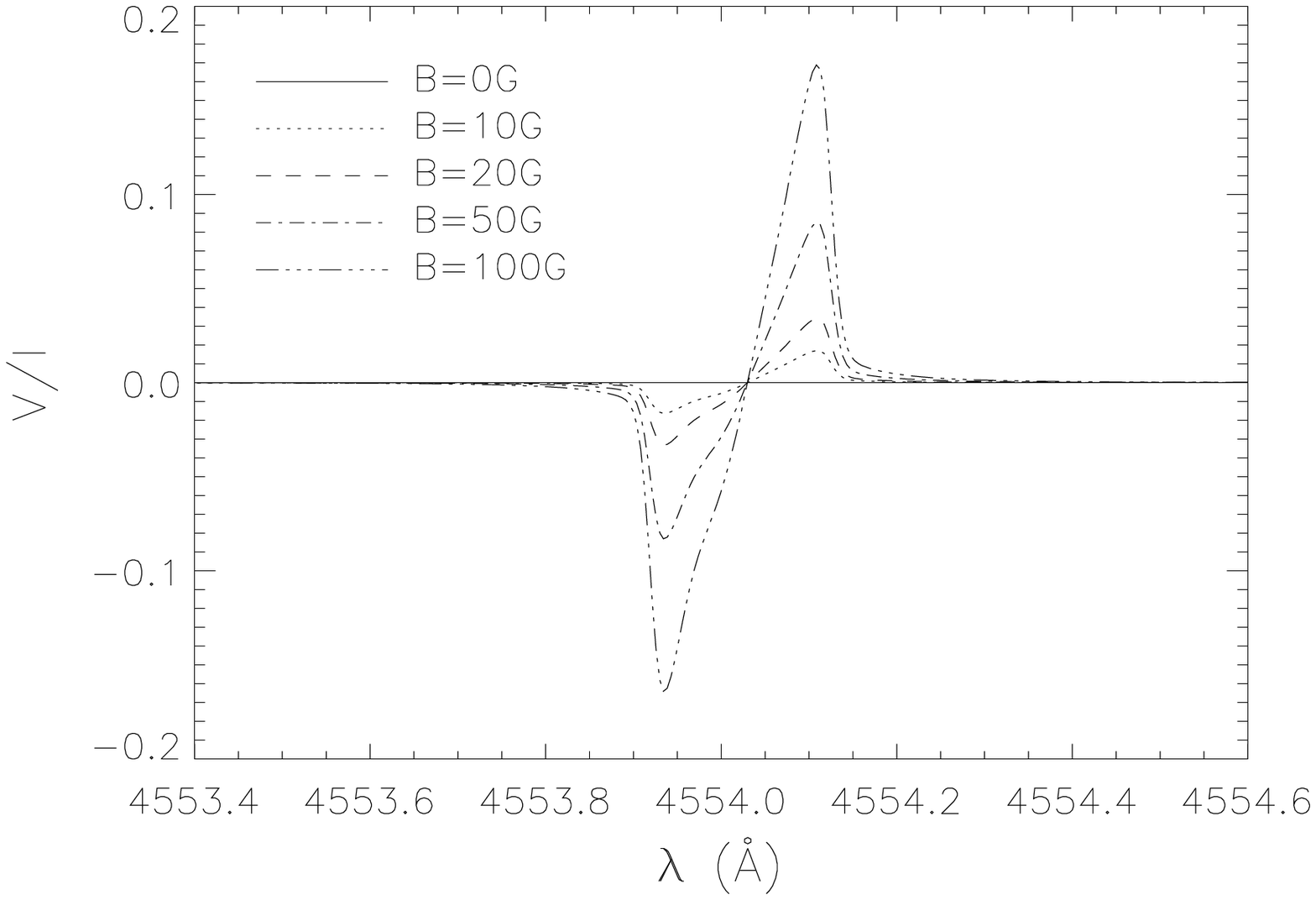}{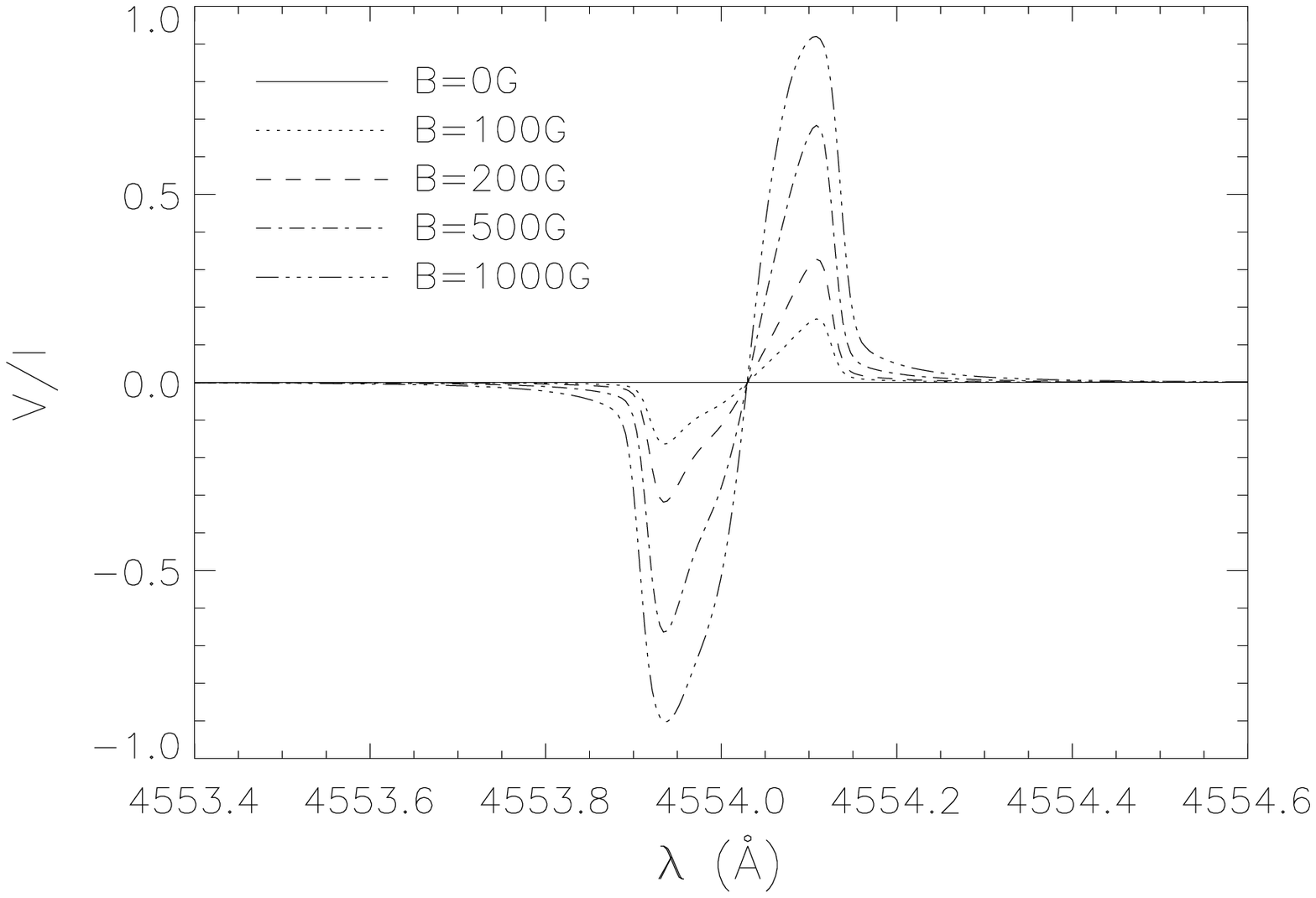}
\caption{{\footnotesize Theoretical $V/I$ profiles of the Ba~{\sc ii} D$_2$ line
in the presence of a longitudinal magnetic field.}}
\label{fig:D2-longitudinal-field-V}
\end{figure}

In the presence of a longitudinal magnetic field of increasing
strength, there is again a decrease of the linear polarization at the
wavelength position of the central peak, due to the Hanle effect, while the
two peaks on the wings are not affected.
In this geometry the Zeeman effect does not modify the linear
polarization and, going to stronger magnetic fields, we enter a regime of
saturation, as shown in the left panel of 
Figure~\ref{fig:D2-longitudinal-field-QU}.
Because of the Hanle effect, we have in this case a rotation of the plane of
linear polarization. This implies the presence of the non zero $U/I$ signal
shown in the right panel of Figure~\ref{fig:D2-longitudinal-field-QU}.
Finally we have a typical antisymmetric $V/I$ signal due to the longitudinal
Zeeman effect, as shown in Figure~\ref{fig:D2-longitudinal-field-V}.
For weak magnetic fields (of the order of about 100~G or
weaker) the signal increases linearly with the magnetic field strength.
Going to stronger fields the linearity is slowly
lost, and the profile starts to saturate.

\subsection{{\it D$_2$ Line -- Random-Azimuth Magnetic Field}}
\label{sect:D2-random-azimuth}
\begin{figure}[!t]
\begin{center}
\includegraphics[width=0.9\textwidth]{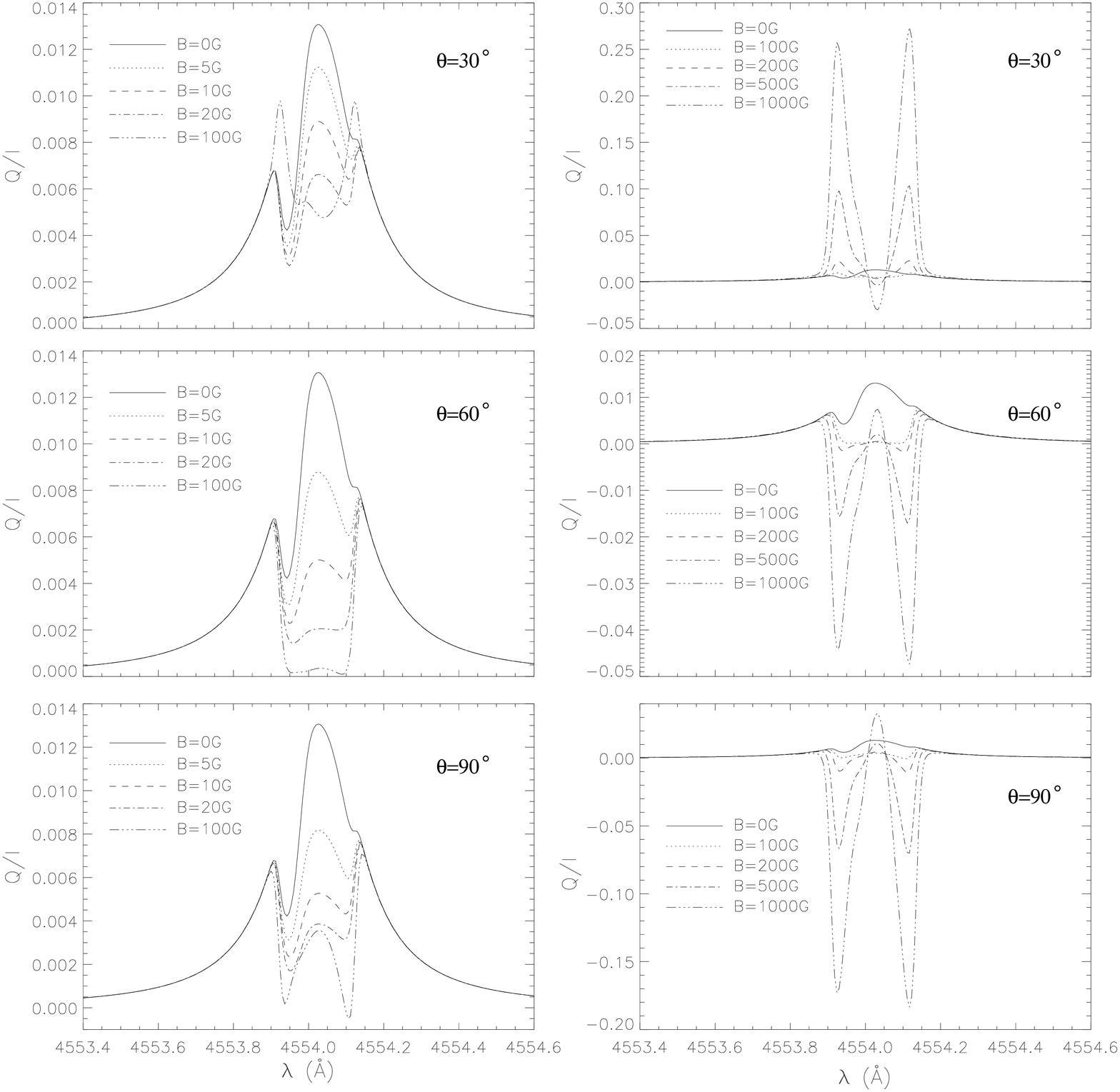}
\caption{{\footnotesize Theoretical $Q/I$ profiles of the Ba~{\sc ii} D$_2$ line
in the presence of a random-azimuth magnetic field for three different 
inclinations: 30$^\circ$ (top), 60$^\circ$ (middle), and 90$^\circ$ (bottom).}}
\label{fig:D2-random-azimuth}
\end{center}
\end{figure}
In this section we present the results obtained for the fractional polarization 
in the presence of magnetic fields with a given inclination and a random 
azimuth (i.e.~the results obtained in the presence of a magnetic field of 
given strength and inclination, averaged over the azimuth). 
Figure~\ref{fig:D2-random-azimuth} shows the theoretical profiles 
obtained in the presence of random-azimuth magnetic fields with inclinations of 
30$^\circ$, 60$^\circ$ and 90$^\circ$.
In the 30$^\circ$ case we observe that increasing the magnetic field 
strength the linear polarization decreases at the wavelength position of 
the central peak and of the two dips close to it, because of the Hanle effect. 
For magnetic fields 
of about 100~G it is possible to observe the first signatures of the Zeeman 
effect, which dominates the linear polarization as we further increase its 
intensity. Note that as the main component of the magnetic field is vertical, 
the Zeeman effect produces the typical three lobes profiles with the same 
signs as in the case of a deterministic vertical magnetic field.
Similar considerations hold for a random-azimuth magnetic field with an 
inclination of 60$^\circ$ and 90$^\circ$. As the main component of the 
magnetic field is now horizontal, as far as 
the Zeeman effect starts to dominate the polarization (which happens for 
fields of about 200~G or stronger in the 60$^\circ$ case, and for fields 
of about 50~G or stronger in the 90$^\circ$ case), we obtain the well known 
Zeeman effect profiles 
with the same signs as in the case of a deterministic horizontal magnetic 
field, perpendicular to the LOS.
Stronger magnetic fields are needed for the Zeeman effect to be 
appreciable in the presence of a random-azimuth magnetic field with an 
inclination of 60$^\circ$ because in this case the vertical and the 
horizontal components of the magnetic field (which produce Zeeman effect 
profiles with opposite signs) are comparable.

\subsection{{\it D$_2$ Line -- Microturbulent Magnetic Field}}
\label{sect:D2-microturbulent}
Averaging the emission coefficient over all the possible orientations of the 
magnetic field\footnote{See Appendix~\ref{app:average} for the details  
implied in performing this average.}, 
we can investigate the polarization properties of the line in the presence 
of a unimodal microturbulent magnetic field.
\begin{figure}[!t]
\begin{center}
\includegraphics[scale=0.32]{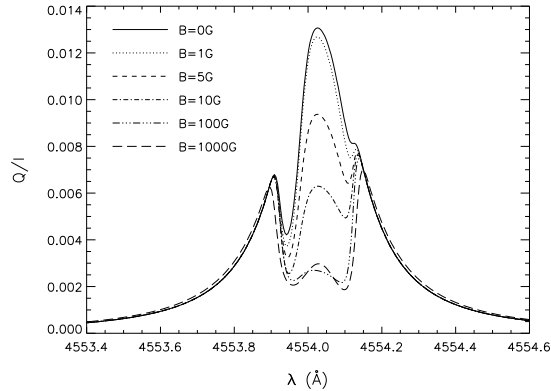}
\caption{{\footnotesize Theoretical $Q/I$ profiles of the Ba~{\sc ii} D$_2$ line
in the presence of a microturbulent magnetic field.}}
\label{fig:D2-microturbulent}
\end{center}
\end{figure}
As shown in Figure~\ref{fig:D2-microturbulent}, we observe that the linear 
polarization at the wavelength position of the central peak decreases,
while the two peaks on the wings remain constant as the magnetic
field strength is increased.
This behaviour can be easily understood: in the presence of a
microturbulent magnetic field there is no observational polarization signal
due to the Zeeman effect, while the Hanle
effect produces a decrease of the linear polarization only in the line-core.
For $B \! > \! 100$~G, we enter a saturation regime and, contrary to the case 
of a longitudinal field, we still have a non-zero signal in the line-core.

\subsection{{\it The Polarization of the D$_{1}$ Line}}
\label{sect:lineD1}
\begin{figure}[!t]
\begin{center}
\includegraphics[angle=270, width=0.95\textwidth]{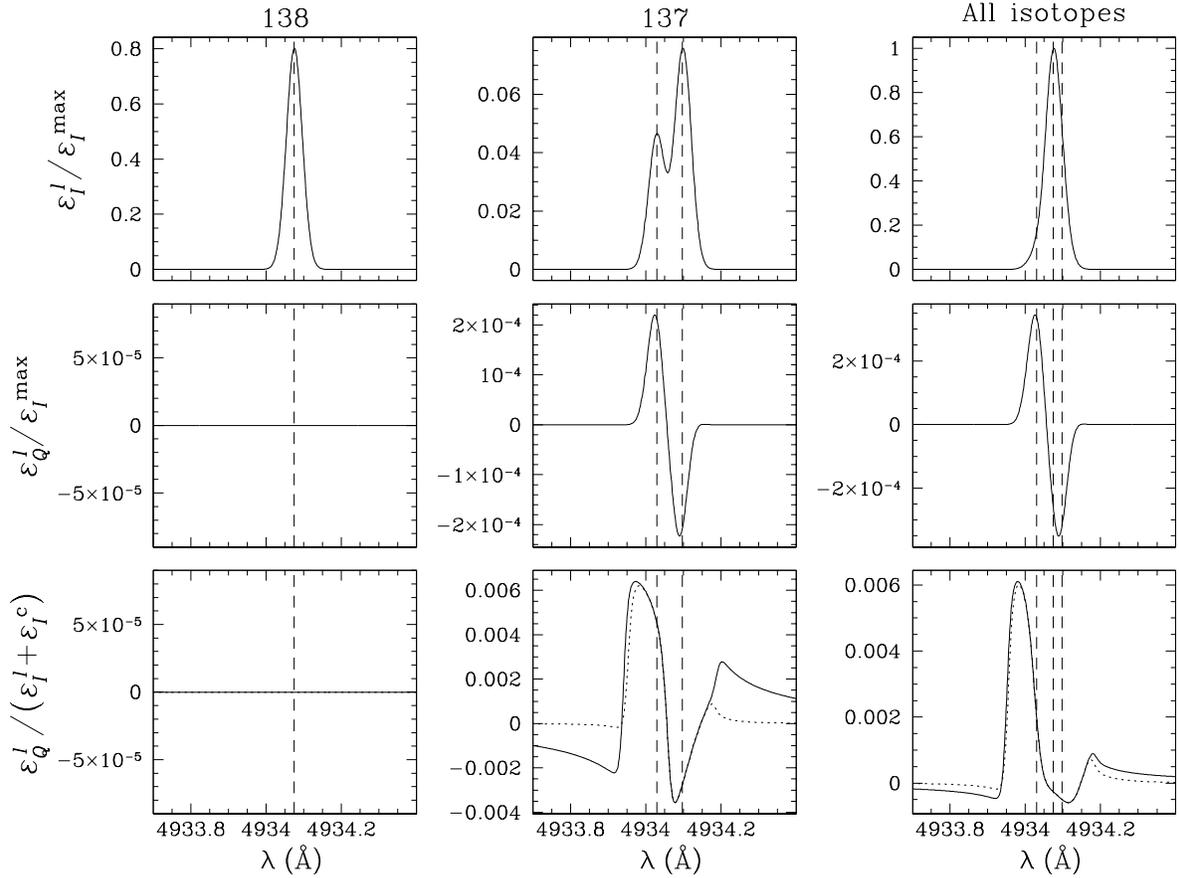}
\caption{{\footnotesize Theoretical profiles of the emission coefficients
$\varepsilon_{I}^{\,l}(\lambda)$ and $\varepsilon_{Q}^{\,l}(\lambda)$,
and of the ratio
$\varepsilon_{Q}^{\,l}(\lambda)/(\varepsilon_{I}^{\,l}(\lambda)+
\varepsilon_{I}^{\, \rm c})$, of the Ba~{\sc ii} D$_1$ line, for the isotope 
138, for the isotope 137 and for all the seven isotopes together, in the 
absence of a magnetic field. The emission profiles are
normalized to the maximum value of the intensity emission profile,
calculated taking into account the contribution of all the seven isotopes,
$(\varepsilon_{I}^{l})_{\rm{max}}$.
The vertical dashed lines show the positions of the three groups
of transitions (see \S~\ref{sect:D2-zero-field}).
The last row shows the
$\varepsilon_{Q}^{\,l}(\lambda)/(\varepsilon_{I}^{\,l}(\lambda)+
\varepsilon_{I}^{\, \rm c})$ profiles without continuum (solid), and
with a continuum $\varepsilon_{I}^{\, \rm c}/(\varepsilon_{I}^{l})_{\rm{max}}$ 
of $9\times 10^{-5}$ (dot).}}
\label{fig:D1-zero-field}
\end{center}
\end{figure}
To understand the physical origin of the polarization signal observed 
in the D$_{1}$ line and, in particular, the role of HFS, we start 
our investigation considering separately the
isotopes 138 (without HFS) and 137 (with HFS), before taking into 
account the contribution coming from all the seven isotopes together.
We use the values of the average number of photons and of the anisotropy factor 
given in \S~\ref{sect:radiation}, and a Doppler width of 30~m{\AA}. 
In Figure~\ref{fig:D1-zero-field} we can see that the emission 
coefficient $\varepsilon_{Q}^{\, l}$ of the isotopes without HFS 
is constant in wavelength and equal to zero, while 
it shows an anti-symmetrical profile, that goes rapidly to zero moving away 
from the line-core, in the isotopes with HFS. For these isotopes the 
frequency integrated Stokes $Q$ emission coefficient is equal to 
zero\footnote{All these properties can be derived analytically, and are 
briefly discussed in Appendix~\ref{app:two-level}.}. 
The profile of the ratio $\varepsilon_{Q}^{\, l}/\varepsilon_{I}^{\,l}$ of 
the isotopes with HFS, contrary to the case of the D$_{2}$ line, goes slowly 
to zero moving away from the line-core.
For this reason the effect of the continuum is less important in this line, as 
it just `pushes' more rapidly to zero the wings of this profile without
modifying significantly its shape.
In the following description of our investigation on the magnetic 
sensitivity of this line we have neglected the contribution to the 
intensity coming from the continuum.\\
\begin{figure}[!t]
\begin{center}
\includegraphics[width=0.95\textwidth]{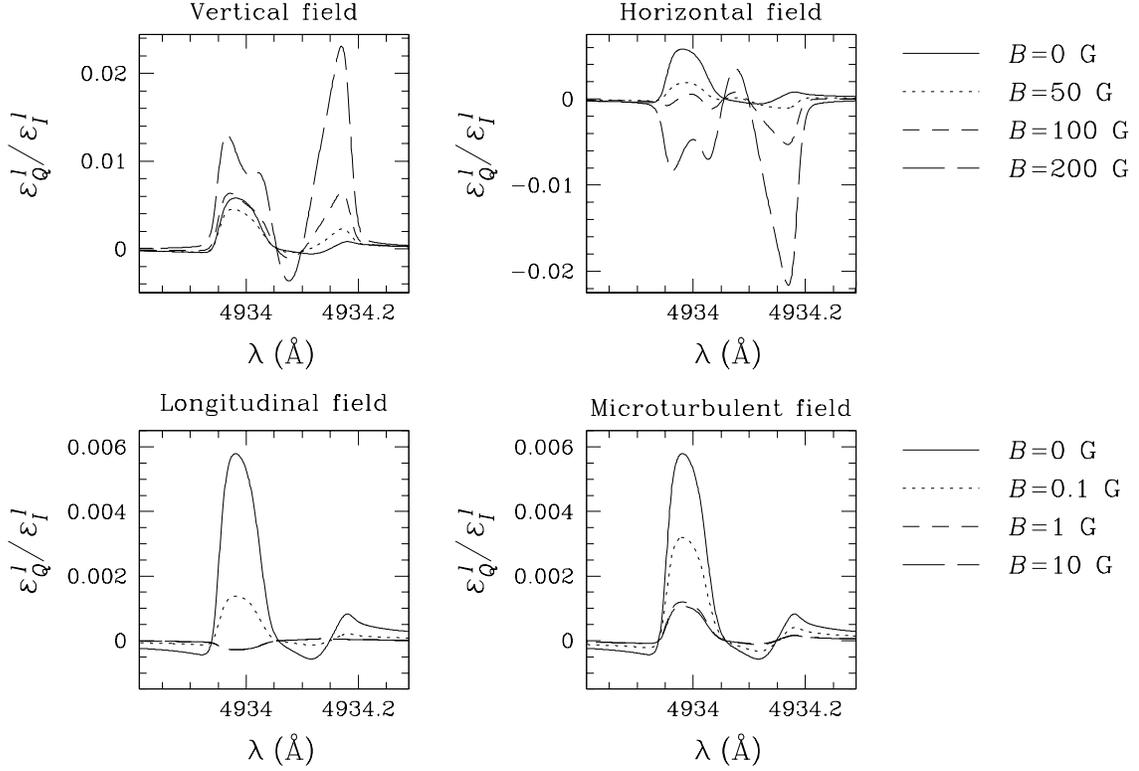}
\caption{{\footnotesize Theoretical profiles for the ratio 
$\varepsilon_{Q}^{\, l}/\varepsilon_{I}^{\, l}$ of the Ba~{\sc ii} D$_1$ line, 
in the presence of a vertical, horizontal perpendicular to the line-of-sight, 
longitudinal and microturbulent magnetic field of various intensities.}}
\label{fig:D1-magnetic-fields}
\end{center}
\end{figure}
\indent As seen in Figure~\ref{fig:D1-zero-field}, within 
the framework of our modeling approach for weak magnetic fields 
($B\! < 50\!$ G) it is not possible to obtain the symmetric 
$Q(\lambda)/I(\lambda)$ profile observed by \citet{Ste00}. The 
theoretical profile that we have obtained has no evident symmetries, and its 
main peak does not coincide in wavelength with the central peak of the 
observed profile. Changing the Doppler width and the anisotropy factor we can 
modify the width and the amplitude of the peaks, adding a continuum 
contribution to the $Q$ Stokes parameter we can shift the profile along the 
polarization scale in order to get values of $Q/I$ in the wings 
close to the observed ones, but we cannot really modify the shape of the profile.
In Figure~\ref{fig:D1-magnetic-fields} we show the theoretical 
$\varepsilon_{Q}^{\, l}/\varepsilon_{I}^{\, l}$ profiles in the presence of a 
vertical, horizontal perpendicular to the line-of-sight, longitudinal and 
microturbulent magnetic field.
In complete analogy with the D$_{2}$ line, the profiles are modified by the 
combined action of the Hanle and Zeeman effects.
As already pointed out by \citet{jtb02} and by \citet{Cas02}
for the Na~{$\!$\sc i} D$_1$ line case, it appears that, at this level of 
approximation, the only way to get a symmetric profile, centered at the 
wavelength position of the central peak of the observed profile 
\citep[see][]{Ste00}, 
is to be in the presence of a magnetic field strong enough to enter the 
transverse Zeeman effect. 
However, as expected, the transverse Zeeman effect produces in the 
$\varepsilon_{Q}^{\, l}/\varepsilon_{I}^{\, l}$ profile two wing lobes that 
are more significant than the central one, which is not the case of the 
observed profile. 

\section{CONCLUSIONS}
\label{sect:conclusions}

The most interesting general conclusion of
our theoretical investigation on the magnetic sensitivity 
of the D-lines of Ba~{$\!$\sc ii} is that
the observation and modeling of the Hanle and Zeeman effects in these resonance lines provide a novel
diagnostic tool for mapping the magnetic fields of the
upper photosphere and lower chromosphere.\\
\indent In particular, the Ba~{$\!$\sc ii} D$_2$ line at 4554~{\AA} is
particularly interesting because the emergent linear polarization has
contributions from different isotopes, contributions that are easily 
resolved and have a different behavior in the presence of a magnetic field. 
As a result, there is a
differential magnetic sensitivity of the emergent linear polarization at 
line center (where the signal is produced by the even isotopes without 
HFS) with respect to the line wings (where the signals are produced by
the odd isotopes with HFS). 
For instance, for the case of a vertical magnetic field
with a strength between 10~G and 100~G, approximately, only the isotopes
with HFS are sensitive to the Hanle effect, which produce an enhancement of
the scattering polarization at the two $Q/I$ wing wavelengths. For the case of
a horizontal field between about 1 and 100~G
the most conspicuous observable effect is the line core depolarization
produced by the Hanle effect of the barium isotopes devoid of HFS.
In both cases, the transverse Zeeman effect begins to play an increasingly
dominating role for field intensities larger than 100~G, approximately.
Useful information on the magnetic sensitivity of the $Q/I$ profile of the
calculated emergent radiation in the D$_2$ line
can be seen in Figure~{\ref{fig:D2-random-azimuth}, which corresponds to the 
case of a random azimuth magnetic field with a fixed inclination.
Of particular interest is the case of an unimodal microturbulent and isotropic 
magnetic field (see Figure~{\ref{fig:D2-microturbulent}}), for which there is 
no contribution from the Zeeman effect and Stokes $U$ and $V$ are zero.\\
\indent Concerning the enigmatic Ba~{$\!$\sc ii} D$_1$ line it is important to
note that in the absence of magnetic fields only the $18\%$ isotopes with
HFS are capable of producing linear polarization through bound-bound
transitions. As with the sodium D$_1$ line, this is possible thanks to the
fact that in the absence of depolarizing mechanisms
only the upper and lower levels of the D$_1$ line 
transition in the odd isotopes are significantly
polarized. Interestingly, a $Q/I$ profile with a conspicuous blue-shifted 
peak is obtained if only the selective emission of
polarization components that results from the upper-level polarization are 
taken into account (see the last panel of the third row of 
Fig.~\ref{fig:D1-zero-field})\footnote{Note, however, that in the absence of a 
magnetic field the theoretical $Q/I^{max}$ profile of the Ba~{$\!$\sc ii} 
D$_1$ line (see the last panel in the second row of 
Fig.~{\ref{fig:D1-zero-field}}) has an antisymmetrical shape, as already found 
for the Na~{$\!$\sc i} D$_1$ line \citep[see Fig.~2 in][]{jtb02}, when only 
selective emission processes are considered.}. 
Under such circumstances one could argue that a detailed radiative
transfer solution for the Ba~{$\!$\sc ii} D$_1$ line including the Doppler
shifts caused by the convective motions and waves that are present in the
solar atmospheric plasma could perhaps produce a symmetric $Q/I$ profile
for the Ba~{$\!$\sc ii} D$_1$ line, as observed by Stenflo et al. (2000).
However, as shown in this paper, for the Ba~{$\!$\sc ii} D$_1$ line we should
expect also a significant contribution from ``zero-field" dichroism --that
is, from the selective absorption of polarization components that results
from the lower-level polarization. In fact, when both selective emission
and absorption processes are taken into account through the approximation
of equation~(\ref{eq:dichroism}), we then obtain the nearly 
antisymmetric $Q/I$ profile of Figure~\ref{fig:D1-dichroism}.
Note that there is no possibility of destroying the lower level polarization
without simultaneously destroying the atomic polarization of the upper level
of the D$_1$ line \citep[see Fig.~1 of][and our Fig.~\ref{fig:sigma20} 
for barium levels]{jtb02}. As far as dichroism is neglected, within the 
framework of our present modeling assumptions one might then be tempted to conclude that the only possibility of obtaining a symmetric $Q/I$ peak for the 
Ba~{\sc ii} D$_1$ line is via the transverse
Zeeman effect, even though, through this kind of mechanism, a profile with wing 
lobes more significant then the central one is obtained, and magnetic fields quite intense are needed. 
Our approach neglects, however, the radiative transfer
effects that we certainly have in the real solar atmosphere, with its
vertical stratification, horizontal inhomogeneities and the Doppler shifts
caused by the above-mentioned upflows, downflows and waves.
Therefore, detailed radiative transfer simulations using realistic solar
atmospheric models are 
urgently needed in order to be able to conclude
whether a symmetric $Q/I$ profile for the Ba~{$\!$\sc ii} D$_1$ line with a significant line-center peak may be
obtained within the framework of the density matrix theory we have applied
in this paper, either because of the influence of the atomic level
polarization of the $18\%$ of the barium isotopes endowed of HFS (which would
require the presence of atmospheric regions with very weak fields), or due
to the transverse Zeeman effect of all the barium isotopes (which would
require the presence of a sufficiently strong magnetic field, and the 
absence of significant saturation effects at the line-center wavelength).\\
\indent Finally, we would like to finish this paper by emphasizing the
importance of pursuing high-spatial resolution polarimetric observations
of the Ba~{$\!$\sc ii} D$_2$ line (e.g., via Fabry-Perot polarimetry) in
order to help decipher the spatial and temporal fluctuations of the magnetic
field vector, in both active and quiet regions of the solar atmosphere.

\acknowledgments
This research has been partially funded by the European Commission through 
the Solar Magnetism Network, and by the Spanish Ministerio de Educaci\'on y 
Ciencia through project AYA2004-05792.

\appendix

\section{RESULTS OBTAINED TAKING INTO ACCOUNT DICHROISM EFFECTS}
\label{app:dichroism}
\begin{figure}[!t]
\begin{center}
\includegraphics[angle=270,width=0.85\textwidth]{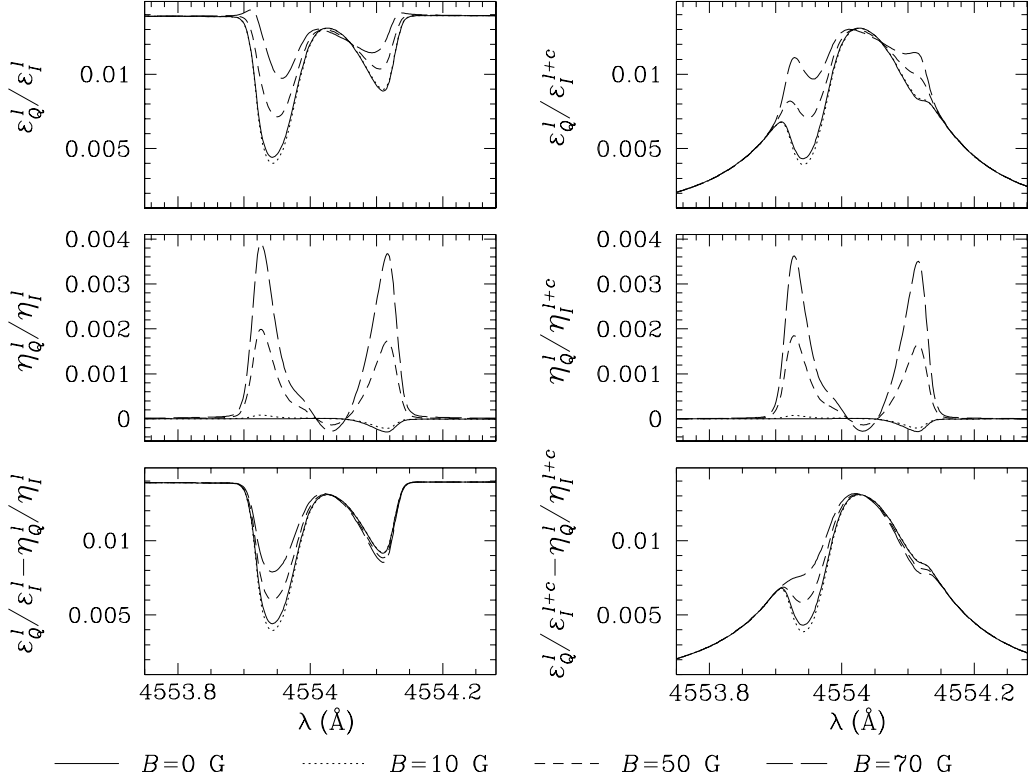}
\caption{{\footnotesize {\it Left column:} theoretical profiles for the 
ratios (from the top) $\varepsilon_{Q}^{\, l}/\varepsilon_{I}^{\, l}$, 
$\eta_{Q}^{\, l}/\eta_{I}^{\, l}$ and for their difference, in the presence 
of a vertical magnetic field of various intensities. {\it Right column:} line 
profiles taking into account the continuum contribution to the intensity 
emission and absorption 
($\varepsilon_{I}^{\, \rm c}/(\varepsilon_{I}^{l})_{\rm{max}}=9 \times 10^{-5}$,
$\eta_{I}^{\, \rm c}$ calculated through eq.~[\ref{eq:eta-cont}]). 
Calculations refer to the Ba~{\sc ii} D$_{2}$ line.}}
\label{fig:D2-dichroismA}
\end{center}
\end{figure}
We show here the results obtained taking into account the absorption effects in 
the D$_{2}$ line. Including the contribution of the continuum, 
equation~(\ref{eq:dichroism}) takes the form
\begin{equation}
\frac{X(\nu,\mathbf{\Omega})}{I(\nu,\mathbf{\Omega})}\approx
\frac{\varepsilon_{X}^{\, l}(\nu,\mathbf{\Omega})}
{\varepsilon_{I}^{\, l}(\nu,\mathbf{\Omega})+
\varepsilon^{\,\rm c}_{I}}-
\frac{\eta_{X}^{\, l}(\nu,\mathbf{\Omega})}
{\eta_{I}^{\, l}(\nu,\mathbf{\Omega})+
\eta^{\,\rm c}_{I}} \;\; ,
\label{eq:dichroism-cont}
\end{equation}
where $\eta_{I}^{\rm c}$ is the continuum contribution to the
total intensity absorption.
This quantity can be calculated from the continuum intensity emission 
coefficient through the relation
\begin{equation}
\eta^{\,\rm c}_{I}=\frac{\varepsilon^{\,\rm c}_{I}}{B_{\rm{P}}
(\nu_{0},T=5800K)} \;\; ,
\label{eq:eta-cont}
\end{equation}
where $B_{\rm{P}}(\nu_{0},T=5800K)$ is the Planck function calculated at the
central wavelength ($\nu_0$) of the line we are investigating and at the
effective solar temperature.
Figures~\ref{fig:D2-dichroismA} and \ref{fig:D2-dichroismB} show the results 
for the D$_{2}$ line. We consider first the results obtained without 
continuum.
\begin{figure}[!t]
\begin{center}
\includegraphics[angle=270,width=0.85\textwidth]{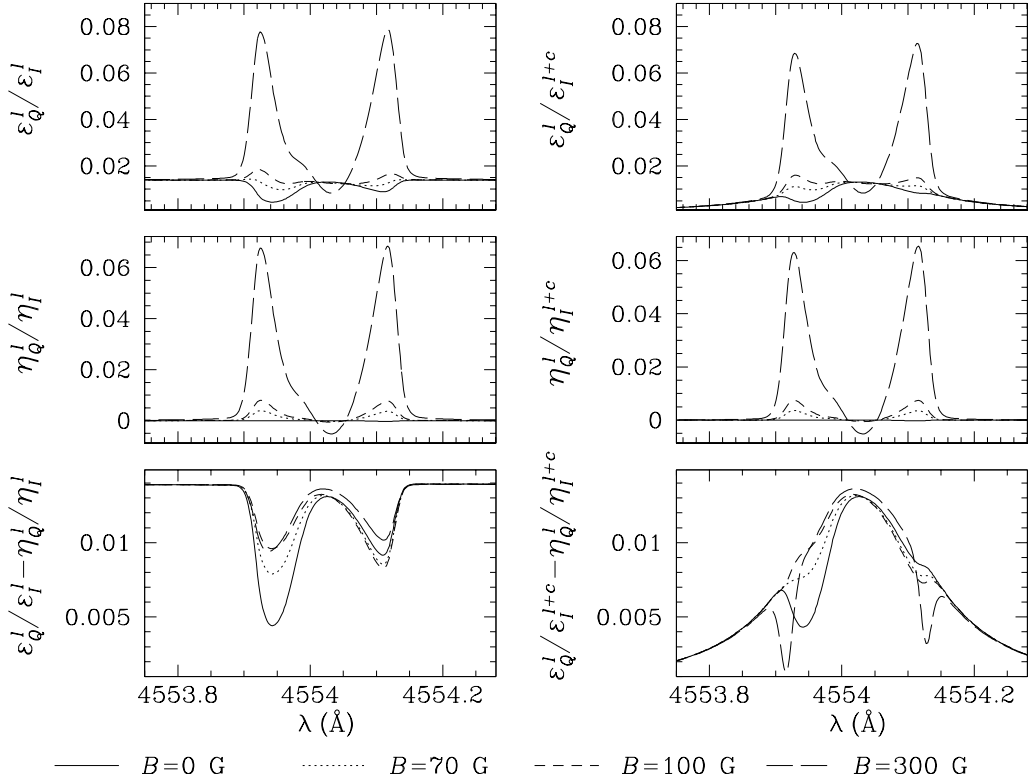}
\caption{{\footnotesize Same as Figure~\ref{fig:D2-dichroismA} for larger 
values of the magnetic field.}} 
\label{fig:D2-dichroismB}
\end{center}
\end{figure}
In the absence of magnetic fields, since the atomic polarization in the lower 
level is much smaller than in the upper level of D$_{2}$ (see 
Fig.~\ref{fig:sigma20}), the absorption effects are completely negligible, as 
already observed in \S~\ref{sect:emergent}. 
Introducing a vertical magnetic field the profile of 
the ratio $\eta_{Q}^{\, l}/\eta_{I}^{\, l}$ is modified by the 
transverse Zeeman effect and, as the magnetic field is increased, it assumes a 
shape which is very similar to the one observed for
$\varepsilon_{Q}^{\, l}/\varepsilon_{I}^{\, l}$ 
(see Fig.~\ref{fig:D2-dichroismB}). For this reason, the profile 
obtained through equation~(\ref{eq:dichroism}) does not show any detail 
due to the transverse Zeeman effect 
since the contribution coming from emission and absorption cancel out. 
The situation is somewhat different in the presence of the continuum. 
In this case, as shown in Figure~\ref{fig:D2-dichroismB}, for magnetic fields 
stronger than about 100~G the contribution coming from the absorption term 
becomes more important. 
At this point it is important to remember that the continuum has been 
considered 
as a parameter in our investigation and, as stressed in 
\S~\ref{sect:D2-zero-field}, we have to be careful when dealing with spectral 
details that find their origin in this physical aspect of the problem.
Similar considerations can be done for the other magnetic field geometries 
considered in this work.\\
\begin{figure}[t]
\begin{center}
\includegraphics[angle=270,width=0.4\textwidth]{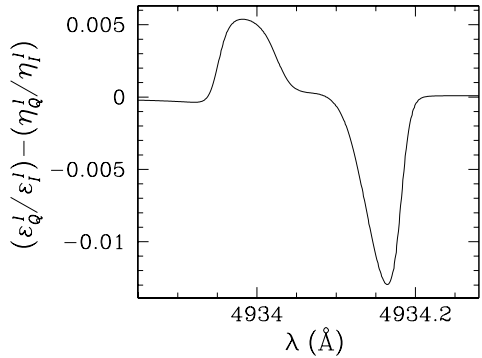}
\caption{{\footnotesize Theoretical $Q/I$ profile of the Ba~{\sc ii} D$_{1}$ 
line calculated according to equation~\ref{eq:dichroism}, in the absence of 
magnetic fields and without any contribution of the continuum.
For the Doppler width, the anisotropy factor, and the average number of photons 
have been used the same values of \S~\ref{sect:lineD1}.}}
\label{fig:D1-dichroism}
\end{center}
\end{figure}
\indent Figure~\ref{fig:D1-dichroism} shows the ratio $(\varepsilon_{Q}^{\, l}/
\varepsilon_{I}^{\, l})-(\eta_{Q}^{\, l}/\eta_{I}^{\, l})$ for the  D$_{1}$ 
line in the absence of magnetic fields. As expected, since the atomic 
polarization is quite similar in the upper and lower levels of the D$_{1}$, 
dichroism is much more important in this line, and the profiles obtained 
through equations~(\ref{eq:fract-polar1}) and (\ref{eq:dichroism}) are quite
different from each other. However, even taking into account dichroism, we 
are not able to reproduce the observed profile.

\section{ANALYTICAL RESULTS FOR A TWO-LEVEL ATOM APPLIED TO THE Ba~{$\!$\sc ii} 
D LINES}
\label{app:two-level}
In this work we have described the Ba~{$\!$\sc ii} ion through a three 
level model atom that allowed us to study both the D$_{1}$ and D$_{2}$ lines; 
we have taken into account the ground level polarization, the stimulated 
emission effects and, finally, the effects of a magnetic field.
At this level a numerical approach of the problem is absolutely necessary.
However, as described in detail in \S~10 of LL04, by introducing some 
simplifying approximations it is possible to obtain analytical expressions 
for the atomic density-matrix elements and, through these, for the radiation 
transfer coefficients of the atomic system. 
These analytical expressions are very useful in order to understand 
the physics of the phenomenon under investigation, which could remain quite 
hidden within a numerical approach.
The basic approximation is to consider a two-level atom. 
The only difference between our three-level model atom and a two-level atom 
lies in the fact that in the SEEs of our model the ground level feels the 
effect of both transitions towards the two upper levels considered. 
However, the numerical solution of the SEEs showed that the upper 
level of the D$_{2}$ line is much more polarized than the ground level,
so that, as far as the D$_{2}$ line is considered, the polarization of the 
ground level, as a first approximation, can be neglected and our atomic 
system can be treated as a two-level atom. 
We can therefore apply the equations of \S~10 of 
LL04 to calculate the upper level density-matrix elements or the 
emission coefficients of the various Ba~{$\!$\sc ii} isotopes, described as 
two-level atoms with HFS, and two-level atoms without HFS, with unpolarized 
lower level.
Neglecting the stimulated emission effects (which is a good approximation 
whenever the incident radiation field, as in our case, is weak), and in 
the absence of the magnetic field, the emission coefficient of a two-level 
atom without HFS is (see eq.[10.16] of LL04)
\begin{eqnarray}
\varepsilon_{i}(\nu,\mathbf{\Omega}) &\! =\! & 
\frac{h\nu}{4\pi}\mathcal{N}_{\ell}
B(\alpha_{\ell}J_{\ell}\rightarrow \alpha_{u}J_{u})\phi(\nu_{0}-\nu) \nonumber\\
& & \times \sum_{KQ}W_{K}(J_{\ell},J_{u})(-1)^{Q}
\mathcal{T}_{Q}^{K}(i,\mathbf{\Omega})J_{-Q}^{K}(\nu_{0}) \;\; ,
\label{eq:eps-two-level}
\end{eqnarray}
where $\mathcal{N}_{\ell}$ is the number density of atoms in the ground 
level\footnote{Note that for a two level atom 
$\mathcal{N}=\mathcal{N}_{\ell}+\mathcal{N}_{u}$ while, for our model atom,
$\mathcal{N}=\mathcal{N}_{\ell}+\mathcal{N}_{u}^{D_{2}}+
\mathcal{N}_{u}^{D_{1}}$. As far as lower level polarization is neglected, 
this is the only difference between a two-level atom and our model atom.}, 
and where 
\begin{equation}
W_{K}(J_{\ell},J_{u})=3(2J_{u}+1)
\Bigg\{ \begin{array}{ccc}
1 & 1 & K \\
J_{u} & J_{u} & J_{\ell}
\end{array} \Bigg\}^{2} \;\; .
\end{equation}
For $90^{\circ}$ scattering of a radiation field with cylindrical 
symmetry around the direction of propagation (as in our case), from 
equation~(\ref{eq:eps-two-level}) we obtain
\begin{equation}
p_{Q}\equiv \frac{\varepsilon_{Q}}{\varepsilon_{I}}=
\frac{3W_{2}(1/2,3/2)}{4/w-W_{2}(1/2,3/2)} \;\; .
\label{eq:fract-pol-two-level}
\end{equation}
This expression shows that for the isotopes without HFS the fractional 
polarization $p_{Q}$ does not depend on frequency, as found in 
\S~\ref{sect:D2-zero-field} (Fig.~\ref{fig:D2-zero-field}).
Substituting the numerical values of the various quantities ($w=0.037$ and 
$W_{2}(1/2,3/2)=0.5$) we obtain the value of 1.4\%, as found with numerical 
calculations.
The expression of the emission coefficient of a two-level atom with HFS 
(eq.~[10.166] of LL04) is much 
more complicated and will not be written here. Anyway, as shown in
\S~10.22 of LL04, at frequencies very distant from the `center 
of gravity', the multiplet behaves in resonance scattering as a simple 
transition between two levels without HFS. The fractional polarization, 
therefore, at these frequencies is still described by 
equation~(\ref{eq:fract-pol-two-level}).
This result justifies the fact that at large distances from the line center 
the fractional polarization of the isotopes with HFS reaches the same value 
as the isotopes without HFS. 
It is important to stress that this asymptotic behaviour of the 
isotopes with HFS is strongly dependent on the interferences between the 
various HFS magnetic sublevels of the D$_{2}$ upper level.
Because of the small frequency distance between the various components of the 
HFS multiplet with respect to their natural width, the analytical expression 
of the emission coefficients of the isotopes with HFS cannot be simplified in 
the neighbourhood of the various transitions. Nevertheless, we can 
qualitatively 
justify the decrease of the fractional polarization at the wavelength positions 
of the various components of the HFS multiplet by considering the frequency 
integrated emission coefficients of the isotopes with HFS
\begin{eqnarray}
\tilde{\varepsilon}_{i}(\mathbf{\Omega}) & \! = \! & \int_{\Delta \nu} 
\varepsilon_i(\nu,\mathbf{\Omega}) {\rm d}\nu \nonumber \\
& = \! & \frac{h\nu}{4\pi}\,\mathcal{N}_{\ell}\,
B(\alpha_{\ell}J_{\ell}\rightarrow \alpha_{u}J_{u})
\sum_{KQ}\,[W_{K}(\alpha_{\ell}J_{\ell}I,\alpha_{u}J_{u})]_{\rm hfs}\,(-1)^{Q}\,
\mathcal{T}_{Q}^{K}(i,\mathbf{\Omega})J_{-Q}^{K}(\nu_{0}) \;\; ,
\end{eqnarray}
where the interval $\Delta \nu$ is sufficiently broad to fully cover all the 
Zeeman components of the line, and where
\begin{equation}
[W_{K}(\alpha_{\ell}J_{\ell}I,\alpha_{u}J_{u})]_{\rm hfs}=W_{K}(J_{\ell},J_{u})
[D_{K}(\alpha_{u}J_{u}I)]_{\rm hfs} \;\; .
\end{equation}
The quantity $[D_{K}(\alpha_{u}J_{u}I)]_{\rm hfs}$ is the depolarizing factor 
due to HFS and it is given by
\begin{eqnarray}
[D_{K}(\alpha JI)]_{\rm hfs} & = & \frac{1}{(2I+1)}\sum_{FF^{\prime}}(2F+1)
(2F^{\prime}+1) 
\Bigg\{ \begin{array}{ccc}
J & J & K \\
F & F^{\prime} & I
\end{array} \Bigg\}^{2} \nonumber \\
& & \times \frac{1}{1+2\pi {\rm i} \nu_{\alpha J I F^{\prime},\alpha J IF}/
A(\alpha J \rightarrow \alpha_{\ell} J_{\ell})} \;\; .
\label{eq:hfs-depolar-fact}
\end{eqnarray}
For the isotope 137 the depolarizing factor 
$[D_{2}(J\!\!=\!\!3/2,I\!\!=\!\!3/2)]_{\rm hfs}$ is equal to 0.27,
and the frequency integrated fractional polarization, that in complete analogy 
with equation~(\ref{eq:fract-pol-two-level}) has the form
\begin{equation}
\tilde{p}_{Q}\equiv\frac{\tilde{\varepsilon}_{Q}}{\tilde{\varepsilon}_{I}}=
\frac{3W_{2}(J_{\ell},J_{u})[D_{K}(\alpha_{u}J_{u}I)]_{hfs}}{4/w-
W_{2}(J_{\ell},J_{u})[D_{K}(\alpha_{u}J_{u}I)]_{hfs}} \;\; ,
\label{eq:hfs-freq-int-fract-pol}
\end{equation}
is equal to 0.0038. Comparing this value with 0.014, the value 
of $p_{Q}$ previously found for the isotopes without HFS (note that for these 
isotopes $p_{Q}$ is equal to $\tilde{p}_{Q}$), we can clearly see 
the depolarizing effect of the HFS.\\
\begin{figure}[!t]
\begin{center}
\includegraphics[angle=270,width=0.4\textwidth]{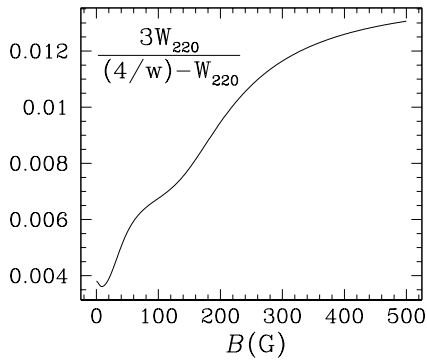}
\caption{{\footnotesize Frequency integrated fractional polarization for the
isotope 137 in a 90$^{\circ}$ scattering event in the presence of a vertical 
magnetic field, as function of the magnetic field strength. Calculations 
refer to the Ba~{\sc ii} D$_2$ line.}}
\label{fig:wkkpqb}
\end{center}
\end{figure}
\indent For the D$_{1}$ line the quantity $W_{2}(1/2,1/2)$ is zero and, 
taking into account equations~(\ref{eq:eps-two-level}) and 
(\ref{eq:hfs-freq-int-fract-pol}), 
it is easy to see that the Stokes $Q$ emission coefficient of the isotopes 
without HFS has to be constant and equal to zero, while the frequency integrated
Stokes $Q$ emission coefficient of the isotopes with HFS has to be zero, as 
found with numerical calculations in \S~\ref{sect:lineD1}.
However, as previously said, for the case of the D$_{1}$ line the approximation 
of unpolarized lower level is not good anymore. Some analytical results that 
it is possible to obtain taking into account the polarization of the lower 
level are derived in LL04.\\
\indent In the presence of a vertical magnetic field 
equation~(\ref{eq:hfs-freq-int-fract-pol}) generalizes into
\begin{equation}
\tilde{p}_{Q}\equiv\frac{\tilde{\varepsilon}_{Q}}{\tilde{\varepsilon}_{I}}=
\frac{3[W_{220}(\alpha_{\ell}J_{\ell}I\alpha_{u}J_{u};B)]_{\rm hfs}}{4/w-
[W_{220}(\alpha_{\ell}J_{\ell}I\alpha_{u}J_{u};B)]_{\rm hfs}} \;\; ,
\label{eq:magnetic-hfs-freq-int-fract-pol}
\end{equation}
where the general expression of the quantity 
$[W_{KK^{\prime}Q}(\alpha_{\ell}J_{\ell}I\alpha_{u}J_{u};B)]_{\rm hfs}$ 
is given in LL04 (eq.~[10.167]).
Figure~\ref{fig:wkkpqb} shows that there is an increase of $\tilde{p}_{Q}$ for 
vertical magnetic fields ranging between 0 and 500~G 
(see Fig.~\ref{fig:D2-vertical-field}).

\section{AVERAGE OF THE EMISSION COEFFICIENTS OVER THE MAGNETIC FIELD 
DIRECTIONS}
\label{app:average}
The only quantity that depends on the magnetic field orientation in the 
general expression of the emission coefficients (\S~\ref{sect:equations}, 
eq.~[\ref{eq:epsilon}]) is the product 
\begin{equation}
\Big[{\mathcal{T}}_{Q}^{K}(j,\mathbf{\Omega})\Big]_{B}
\Big[{}^{\alpha_{u}J_{u}I}\!\rho_{Q_{u}}^{K_{u}}(F_{u}^{\prime},F_{u}^
{\prime\prime})\Big]_{B} \; \; ,
\label{eq:magnetic-ref}
\end{equation}
where the label $B$ means that the quantity is calculated in the 
magnetic field reference system.
We already observed (\S~\ref{sect:polarization}) that the spherical statistical
tensors calculated in the magnetic reference system, because of the symmetry
of the problem, do not depend on the azimuth of the magnetic field, $\chi_{B}$,
but only on its inclination with respect to the local vertical, $\theta_{B}$.
For this reason all the dependence on $\chi_{B}$ is included into the 
geometrical tensor $\big[{\mathcal{T}}_{Q}^{K}(j,\mathbf{\Omega})\big]_{B}$.
The expression of the tensor $\mathcal{T}_{Q}^{K}$ in terms of rotation 
matrices is given by (eq.~[5.159] of LL04)
\begin{equation}
\Big[{\mathcal{T}}_{Q}^{K}(j,\mathbf{\Omega})\Big]_{B}=
\sum_{P}t^{K}_{P}(j)\mathcal{D}^{K}_{PQ}(R_0) \;\; ,
\label{eq:tauKQ}
\end{equation}
where $\mathcal{D}$ is the rotation matrix, $R_0$ is the rotation bringing 
the reference system $(\mathbf{e}_{\it{a}}(\mathbf{\Omega}),\mathbf{e}_{\it{b}}
(\mathbf{\Omega}),\mathbf{\Omega})$, with 
$\mathbf{e}_{\it{a}}(\mathbf{\Omega})$ the 
reference direction, into the reference system with the $z$ axis 
directed along the magnetic field, and where $t^{K}_{P}(j)$ is a scalar 
quantity that does not depend on the particular geometry of the problem 
(cf.~eq.~[5.160] of LL04).
The relation between the geometrical tensor 
$\mathcal{T}_{Q}^{K}$ calculated in the magnetic field reference system, and 
the same quantity calculated in the local vertical reference system 
(with the $z$ axis directed along the local vertical) is
\begin{equation}
\Big[{\mathcal{T}}_{Q}^{K}(j,\mathbf{\Omega})\Big]_{B}=
\sum_{A}\Big[{\mathcal{T}}_{A}^{K}(j,\mathbf{\Omega})\Big]_{V}
\mathcal{D}^{K}_{AQ}(R_1) \;\; ,
\label{eq:tauKQ-rotat}
\end{equation}
where the label $V$ means that the corresponding quantity is calculated into
the local vertical reference system, and
where $R_1$ is the rotation bringing the local vertical reference system into 
the magnetic field reference system. In the geometry of our problem the 
rotation $R_1$ is defined by the Euler angles $(\chi_B,\theta_B,0)$.
Obviously, because of its definition, 
$\big[{\mathcal{T}}_{A}^{K}(j,\mathbf{\Omega})\big]_{V}$ does not 
depend on the magnetic field orientation: in 
analogy with equation~(\ref{eq:tauKQ}) we can write
\begin{equation}
\Big[{\mathcal{T}}_{A}^{K}(j,\mathbf{\Omega})\Big]_{V}=
\sum_{P}t^{K}_{P}(j)\mathcal{D}^{K}_{PA}(R_2) \;\; ,
\end{equation}
where $R_2$ is the rotation that brings the reference system 
$(\mathbf{e}_{\it{a}}(\mathbf{\Omega}),\mathbf{e}_{\it{b}}(\mathbf{\Omega}),
\mathbf{\Omega})$ into the local vertical reference system.
Substituting equation~(\ref{eq:tauKQ-rotat}) into the expression 
(\ref{eq:magnetic-ref}) we obtain
\begin{equation}
\Big[{\mathcal{T}}_{Q}^{K}(j,\mathbf{\Omega})\Big]_{B}
\Big[{}^{\alpha_{u}J_{u}I}\!\rho_{Q_{u}}^{K_{u}}(F_{u}^{\prime},F_{u}^
{\prime\prime})\Big]_{B}=\sum_{A}\Big[{\mathcal{T}}_{A}^{K}
(j,\mathbf{\Omega})\Big]_{V}\mathcal{D}^{K}_{AQ}(R_1)
\Big[{}^{\alpha_{u}J_{u}I}\!\rho_{Q_{u}}^{K_{u}}(F_{u}^{\prime},F_{u}^
{\prime\prime})\Big]_{B} \;\; ,
\end{equation}
where we see that all the dependence on $\chi_B$ is included in the rotation
matrix $\mathcal{D}^{K}_{AQ}(R_1)$.
Recalling the expression of the rotation matrices in terms of the reduced 
rotation matrices (cf. eq.~[2.68] of LL04), and recalling the Euler 
angles corresponding to the rotation $R_1$, we have
\begin{equation}
\mathcal{D}^{K}_{AQ}(R_1)={\rm{e}}^{-{\rm{i}}A\chi_{B}}d^{K}_{AQ}(\theta_{B})
\;\; ,
\end{equation}
where $d^{K}_{AQ}$ is the reduced rotation matrix, which
depends only on the second Euler angle of the rotation, in our 
case the inclination of the magnetic field.
Averaging on $\chi_{B}$ reduces therefore to calculate the integral
\begin{equation}
\frac{1}{2\pi}\int_{0}^{2\pi}{\rm{e}}^{-{\rm{i}}A\chi_{B}}{\rm{d}}\chi_{B} \;\;.
\end{equation}
It is easy to see that the integral is different from zero only if $A=0$, and 
that, in this case, it is equal to 1.
Averaging the emission coefficients over $\chi_{B}$, the magnetic field 
azimuth, is therefore equivalent to substitute expression 
(\ref{eq:magnetic-ref}) with
\begin{equation}
\Big[{\mathcal{T}}_{0}^{K}(j,\mathbf{\Omega})\Big]_{V}
d^{K}_{0Q}(\theta_B)
\Big[{}^{\alpha_{u}J_{u}I}\!\rho_{Q_{u}}^{K_{u}}(F_{u}^{\prime},F_{u}^
{\prime\prime})\Big]_{B} \;\; .
\end{equation}
Now all the quantities depend only on the magnetic field inclination, and we 
can complete numerically (for example by a Gaussian quadrature)
the average over the magnetic field orientation.


\begin{thebibliography}{}
\bibitem[Becker et al.(1981)]{Bec81}
Becker, W., Blatt, R., \& Werth, G. 1981, J. Physique Coll., 42, 
C8-339
\bibitem[Bommier(1980)]{Bom80}
Bommier, V. 1980, \aap, 87, 109
\bibitem[Casini et al.(2002)]{Cas02}
Casini, R., Landi Degl'Innocenti, E., Landolfi, M., \& Trujillo Bueno, J. 
2002, \apj, 573, 864
\bibitem[Landi Degl'Innocenti \& Landolfi(2004)]{Libro}
Landi Degl'Innocenti, E., \& Landolfi, M. 2004, Polarization in Spectral
Lines (Dordrecht: Kluwer)
\bibitem[Kopfermann(1958)]{Kop58}
Kopfermann, H. 1958, Nuclear Moments (New York: Academic Press)
\bibitem[Moore(1958)]{Moore}
Moore, C. E. 1958, Atomic Energy Levels: as derived from the analyses of 
optical spectra, Vol.III (Washington National Bureau of Standards)
\bibitem[Pierce(2000)]{Allen}
Pierce, K. 2000, in Allen's Astrophysical Quantities, ed. A. N. Cox
(4th ed.; New York: Springer), 355 
\bibitem[Stenflo(1997)]{Ste97b}
Stenflo, J. O. 1997, \aap, 324, 344
\bibitem[Stenflo \& Keller(1997)]{Ste97a}
Stenflo, J. O., \& Keller, C. U. 1997, \aap, 321, 927
\bibitem[Stenflo et al.(1998)]{Ste98}
Stenflo, J. O., Keller, C. U., \& Gandorfer, A. 1998, \aap, 329, 319
\bibitem[Stenflo et al.(2000)]{Ste00}
Stenflo, J. O., Keller, C. U., \& Gandorfer, A. 2000, \aap, 335, 789
\bibitem[Stenflo et al.(2002)]{Ste02}
Stenflo, J. O., Gandorfer, A., Holzreuter, R., Gisler, D., Keller, C. U.,
\& Bianda, M. 2002, \aap, 389, 314
\bibitem[Stenflo(2003)]{Ste03}
Stenflo, J.O. 2003, in ASP Conf. Ser. 307, Solar Polarization 3, 
ed. J. Trujillo Bueno, \& J. S\'anchez Almeida (San Francisco: ASP), 385
\bibitem[Trujillo Bueno(2003)]{jtb03}
Trujillo Bueno, J. 2003, in ASP Conf. Ser. 307, Solar Polarization 3,
ed. J. Trujillo Bueno, \& J. S\'anchez Almeida (San Francisco: ASP), 407
\bibitem[Trujillo Bueno et al.(2002)]{jtb02}
Trujillo Bueno, J., Casini, R., Landolfi, M., \& Landi Degl'Innocenti, E. 
2002, \apj, 566, L53
\bibitem[Villemoes et al.(1993)]{Vil93}
Villemoes, P., Arnesen, A., Heijkenskj\"old, F., \& W\"annstr\"om, A. 1993,
J. Phys. B, 26, 4289
\bibitem[Wendt et al.(1984)]{Wen84}
Wendt, K., Ahmad, S. A., Buchinger, F., Mueller, A. C., Neugart, R., \&
Otten, E. W. 1984, Z. Phys. A, 318, 125
\bibitem[Wendt et al.(1988)]{Wen88}
Wendt, K., Ahmad, S. A., Ekstr\"om, C., Klempt, W., Neugart, R., \& Otten, 
E. W. 1988, Z. Phys. A, 329, 407
\end{thebibliography}
\end{document}